\documentclass{aastex63}

\usepackage{sidecap}

\usepackage{tabularx, amsmath, amssymb}
\usepackage{multirow}

\newcommand{\HI}{H{\sc ~i}}
\newcommand{\HeI}{He{\sc ~i}}
\newcommand{\HeII}{He{\sc ~ii}}

\newcommand{\CII}{C{\sc ~ii}}
\newcommand{\CIII}{C{\sc ~iii}}

\newcommand{\NII}{N{\sc ~ii}}
\newcommand{\NIII}{N{\sc ~iii}}
\newcommand{\NV}{N{\sc ~v}}
\newcommand{\NIV}{N{\sc ~iv}}
\newcommand{\CIV}{C{\sc ~iv}}
\newcommand{\OI}{O{\sc ~i}}
\newcommand{\OII}{O{\sc ~ii}}
\newcommand{\OIII}{O{\sc ~iii}}
\newcommand{\OIV}{O{\sc ~iv}}
\newcommand{\OVI}{O{\sc ~vi}}
\newcommand{\OV}{O{\sc ~v}}
\newcommand{\OVII}{O{\sc ~vii}}
\newcommand{\OVIII}{O{\sc ~viii}}
\newcommand{\SiIII}{Si{\sc ~iii}}
\newcommand{\SiIV}{Si{\sc ~iv}}
\newcommand{\MgII}{Mg{\sc ~ii}}
\newcommand{\cmminusthree}{\,cm$^{-3}$\;}

\newcommand{\Zsun}{Z_{\rm \odot}}

\def\cmc{\ifmmode {\rm \: cm^{-2}} \else $\rm \: cm^{-2}$\fi}
\def\cmv{\ifmmode {\rm \: cm^{-3}} \else $\rm \: cm^{-3}$\fi}
\def\fphosr{\ifmmode {\rm \: \gamma~cm^{-2}~s^{-1}~sr^{-1}} \else $\rm \: \gamma~cm$^{-2}$~s$^{-1}$~sr^{-1}$\fi}
\def\fergsr{\ifmmode {\rm \: erg~cm^{-2}~s^{-1}~sr^{-1}} \else $\rm \: erg~cm$^{-2}$~s$^{-1}$~sr$^{-1}$\fi}
\def\fergsec{\ifmmode {\rm \: erg~cm^{-2}~s^{-1}~arcsec^{-2}} \else $\rm \: erg~cm$^{-2}$~s$^{-1}$~arcsec$^{-2}$\fi}
\def\fphosec{\ifmmode {\rm \: \gamma~cm^{-2}~s^{-1}~arcsec^{-2}} \else $\rm \: \gamma~cm$^{-2}$~s$^{-1}$~arcsec$^{-2}$\fi}

\begin{document}

\renewcommand{\thefootnote}{\fnsymbol{footnote}}

\title{Absorption-based Circumgalactic Medium Line Emission Estimates}
\author{Daniel R. Piacitelli
$^*$}\footnote{piacitelli.danielr@gmail.com}
\affiliation{Astronomy Department, University of Washington, Seattle, WA 98195, USA}
\author{Erik Solhaug
$^{\dagger}$} \footnote{ersolhaug@gmail.com}
\affiliation{Astronomy Department, University of Washington, Seattle, WA 98195, USA}
\author{Yakov Faerman}
\affiliation{Astronomy Department, University of Washington, Seattle, WA 98195, USA}
\author{Matthew McQuinn}
\affiliation{Astronomy Department, University of Washington, Seattle, WA 98195, USA}

\renewcommand{\thefootnote}{\arabic{footnote}}

\shorttitle{Estimated CGM Emission}

\begin{abstract}
Motivated by integral field units (IFUs) on large ground telescopes and proposals for ultraviolet-sensitive space telescopes to probe circumgalactic medium (CGM) emission, we survey the most promising emission lines and how such observations can inform our understanding of the CGM and its relation to galaxy formation. We tie our emission estimates to both HST/COS absorption measurements of ions around $z\approx 0.2$ Milky Way mass halos and models for the density and temperature of gas. We also provide formulas that simplify extending our estimates to other samples and physical scenarios. We find that \OIII~5007\,\AA\ and \NII~6583\,\AA, which at fixed ionic column density are primarily sensitive to the thermal pressure of the gas they inhabit, may be detectable with KCWI and especially IFUs on 30\,m telescopes out to half a virial radius.  \OV~630\,\AA\ and \OVI\ $1032,1038$\,\AA\ are perhaps the most promising ultraviolet lines, with models  predicting intensities $>100~\gamma$~cm$^{-2}$~s$^{-1}$~sr$^{-1}$ in the inner 100\;kpc of Milky Way-like systems. A detection of \OVI\ would confirm the collisionally ionized picture and constrain the density profile of the CGM. Other ultraviolet metal lines constrain the amount of gas that is actively cooling and mixing. We find that \CIII\, 978\,\AA\ and \CIV\, 1548\,\AA\ may be detectable if an appreciable fraction of the observed \OVI\ column is associated with mixing or cooling gas. H$\alpha$ emission within $100\;$kpc of Milky Way-like galaxies is within reach of current IFUs even for the minimum signal from ionizing background fluorescence, while Hydrogen $n>2$ Lyman-series lines are too weak to be detectable.
\end{abstract}

\section{Introduction}

The low redshift circumgalactic medium (CGM) -- the medium that extends $\gtrsim$ 200 kpc around galaxies -- is thought to harbor a large fraction of the cosmic baryons. However, the amount of gas that resides within the CGM, and especially how this gas is cooling, mixing and flowing, is not well understood \citep[e.g.][]{2017ARA&A..55..389T}. Understanding the CGM is key to understanding galaxy formation, as the CGM is the trough from which galaxies must feed to grow.  Absorption spectroscopy is the most sensitive probe of low column density material in the extended CGM. For the low-z CGM, absorption studies came into full force in the last decade or so with the Cosmic Origins Spectrograph (COS) on the Hubble Space Telescope \citep[HST;][]{tumlinson11, 2014ApJ...791..128S, werk16}. Still, despite UV spectra providing sightlines through a significant sample of CGMs, the community has not reached a consensus on what these absorption observations mean for the gas distribution and dynamics \citep[e.g.][]{2017ARA&A..55..389T}. Thus, new observables are likely required to make progress.

One frontier is to observe the CGM in emission. Past CGM emission work has mainly targeted quasar environments where the larger photoionizing flux can greatly enhance emission \citep[e.g.][]{2021MNRAS.503.3044F}. There are few constraints at low redshifts in non-AGN galactic  environments \citep[e.g.][]{2011ApJ...728...55R, 2017MNRAS.467.4802F,2018ApJ...861...34Z, 2019ApJ...880...28Z}, with much of the focus on the Milky Way itself -- especially 21cm, H$\alpha$, and \OVII\ and \OVIII\ lines \citep[e.g.][]{putman03, putman12, bregman07, 2015ApJ...800...14M}. There is also a burgeoning focus on observing CGM optical line emission around compact starburst galaxies using the newest integral field units (IFUs), finding some systems that show surprisingly large spatial extents for this emission \citep{2019ApJ...882...17Y, 2019Natur.574..643R, 2021ApJ...909..151B, 2021MNRAS.507.4294Z}, observations which compliment past ultraviolet \OVI\ measurements in the inner $10\,$kpc of a few star forming galaxies \citep{2003ApJ...591..821O, hayes}. Additionally, there are many instruments being proposed and even coming online that may be capable of observing the CGM of external galaxies in emission: from IFUs on large terrestrial telescopes (MUSE, KCWI, and eventually these on the 30~meters; \citealt{2010SPIE.7735E..08B, 2012SPIE.8446E..13M}) and imagers on small (Dragonfly; \citealt{2014PASP..126...55A}), to UV-sensitive balloons and satellites  (FIREBall, Aspera; potentially CETUS, Maratus, HALO, and a 6~meter recommended by the 2020 decadal survey; \citealt{2014SPIE.9144E..30G, aspera, cetus}) and X-ray telescopes (HUBS, ATHENA, LYNX; \citealt{hubs, 2020AN....341..224B,Lynx1}). 

Theoretical work on the CGM has focused primarily on absorption.  The number of studies predicting CGM line emission is more limited  \citep{2013MNRAS.430.2688V, 2016ApJ...827..148C, 2016MNRAS.463..120S, 2019ApJ...877....4L,2019MNRAS.489.2417A, faerman20, corlies20,2021MNRAS.506.5129B,2021MNRAS.507.4445N}. The emission picture is further complicated by the fact that galaxy formation simulations -- the resource used to predict emission in the vast majority of studies -- generally do not match the observed ionic column densities \citep[e.g.][]{2017ARA&A..55..389T}, although recent simulations have more successfully reproduced the observed columns in higher ions especially \OVI\ \citep[e.g.][]{nelson18}.

The aim of this paper is to make predictions for the potential intensities of CGM emission in ultraviolet and optical emission lines. Rather than directly using cosmological simulations, we instead base our emission estimates on the empirical results of absorption line studies, in particular the measured ionic column densities and, in some instances, constraints on the gas density from equilibrium ionization models. An advantage of our approach is that it provides an understanding of which intensities may be possibly tied to the ionic column density -- the physical property of the CGM that is often the most observationally constrained. We also provide simple formulae and collisional cross sections that allow others to easily reproduce and extend our calculations.

We place some restrictions on the scope of this study. We do not concentrate on ions that are most likely to be present when the hydrogen is neutral such as \OI, \CII\ and \MgII.  Emission in these iconic ions is most sensitive to their photoionization rate and total column, and less sensitive to the CGM density and temperature.  Secondly, we do not consider X-ray lines, such as \OVII\ and \OVIII.  At present there are only a few absorption detections in extragalactic systems, preventing our method of analysis. Additionally, these X-ray lines are most sensitive to virial temperature gas for Milky Way-like systems, whereas the ions we consider probe cooler phases.

This paper is organized as follows. Section~\ref{sec:physics} provides a lightning overview of the most promising emission lines and their expected intensities. Then, Section~\ref{sec:results} discusses the emission from various lines, where each subsection concentrates on a different class (\OVI, optical lines, all potentially important UV lines, and lastly \HI). We do not use the traditional bracket notation to differentiate forbidden transitions. Unless indicated otherwise, plots show the full COS-Halos sample of CGM column densities for a given ion. This sample targets $L_*$ galaxies, including both star-forming and quiescent galaxies.  For reference, while we use surface brightness units of $\gamma$~cm$^{-2}$ s$^{-1}$~sr$^{-1}$, to convert to another commonly used unit use 1 $\gamma$ cm$^{-2}$ s$^{-1}$~sr$^{-1}$ = $3.9\times10^{-22}\, (10^3{\rm \,\AA}/\lambda)\,(1.2/[1+z])$~erg~cm$^{-2}$~s$^{-1}$ arcsec$^{-2}$, where $\lambda$ is the emitted wavelength. \\ 

\section{the physics of CGM emission}
\label{sec:physics}

This section provides simple estimates for the intensities in CGM emission lines and the relevant physics, with \S~\ref{sec:results} considering the most promising lines in more detail.

\subsection{Energetic Considerations}
\label{sec:energetics}

\begin{figure}
\includegraphics[width=18.5cm]{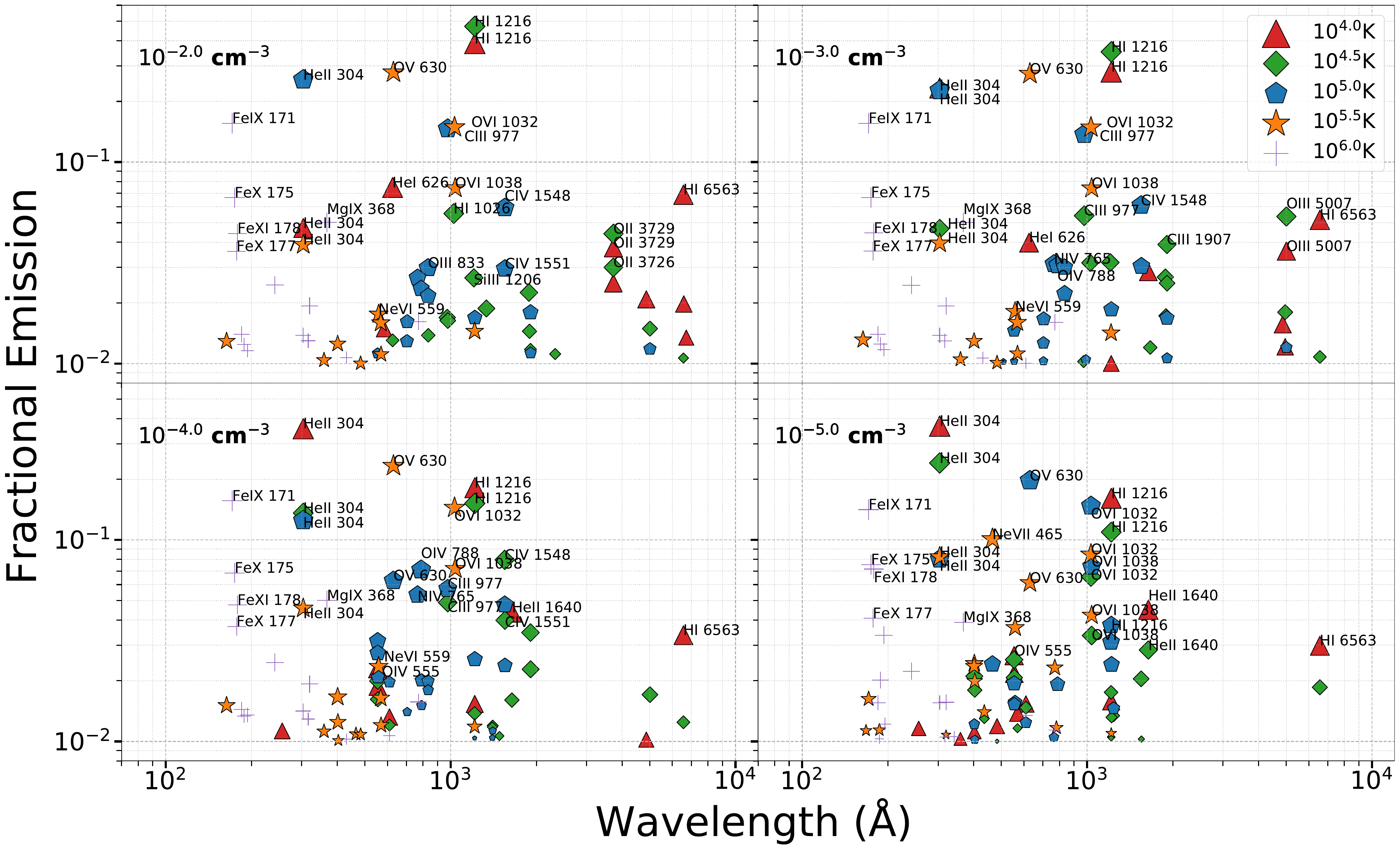}
\caption{Line emissivity as a fraction of the total emissivity for the gas ($\alpha_{\lambda_X}$ in eqn.~\ref{eqn:Ienergy}), where the $x$-axis shows the wavelength of the transition $\lambda_X$, and the legend specifies the assumed gas temperature for each marker type. Each panel shows the \{$\lambda_X$, $\alpha_{\lambda_X}$\} for different total hydrogen densities of $n=10^{-2}$ (top left),~$10^{-3}$ (top right), $10^{-4}$ (bottom left) and $10^{-5}$cm$^{-3}$ (bottom right). These calculations assume the \emph{Q18} UV background model from \citet{KS19}, and a metallicity of $Z'=0.3$~solar. For each temperature considered, the points represent the most intense lines with fractional emission greater than $1\%$ of the total, and the five largest line contributors are annotated rightward of the corresponding point.  A larger marker size indicates a greater fractional emission.  The data for each line, as well as lines with fractional emission below $1\%$, is included in the online submission of this paper.}
\label{fig:f1}
\end{figure}

We start with a simple estimate for the anticipated emission strengths in CGM lines, assuming they are powered by galactic feedback processes. Let us assume that a fraction $f_{\rm CGM}$ of the energy in feedback results in cooling emission from the CGM. If the CGM emits over a radial extent of $R_{\rm CGM}$ with a fraction $\alpha_{\lambda_X}$ of the energy coming out in transition $X$ with wavelength $\lambda_X$, this yields an average observed intensity emitted over an area of $\pi R_{\rm CGM}^2$ of
\begin{equation}
     I_{\lambda_X} = \frac{\alpha_{\lambda_X} f_{\rm CGM} P_{\rm SN}}{4\pi^2 h\nu_{\rm X}  R_{\rm CGM}^2 (1+z)^3} =  290 ~\gamma {\rm ~cm}^{-2} ~{\rm s}^{-1} ~{\rm sr}^{-1}   f_{\rm CGM} \left(\frac{\alpha_{\lambda_X}}{0.03}\right)  \left(\frac{\rm SFR}{5 M_\odot {\rm ~yr}^{-1}}\right) \left(\frac{R_{\rm CGM}}{100\; {\rm kpc}}\right)^{-2}  \left(\frac{\lambda_X}{1000\, {\rm \AA}}\right)^{-1}   \left(\frac{1+z}{1.3}\right)^{-3},
     \label{eqn:Ienergy}
\end{equation}
where $P_{\rm SN}$ is the power in supernova feedback.\footnote{Here we assume that supernovae dominate the feedback energetics, but this calculation can be easily generalized to include black hole feedback with a certain power.} For the rightmost equality, we take $P_{\rm SN}$  to be $10^{51}$~erg per supernovae times 1\% of the star formation rate (SFR), motivated by estimates for the kinetic energy per supernovae and the fraction of stellar mass formed per supernovae. We note that several works suggest $ f_{\rm CGM}\sim 0.3-1$ \citep{tumlinson11, fielding17, faerman17, mcquinn18}, and emission observations could test such predictions. 
For $\alpha_{\lambda_X} \gtrsim 10^{-2}-10^{-3}$, the average intensity given by equation~(\ref{eqn:Ienergy}) may be detectable. Existing IFUs on $\sim 10$m optical telescopes are sensitive to surface brightness as low as $I_{\lambda_X} \sim 300-1000 ~\gamma {\rm ~cm}^{-2} ~{\rm s}^{-1} ~{\rm sr}^{-1}$ \citep{2010SPIE.7735E..0MM, 2012SPIE.8446E..13M, 2019ApJ...877....4L}, with IFUs on future extremely large telescopes projecting factor of $\sim 3-10$ improvements \citep{2019ApJ...877....4L}.  In the UV, a hypothetical Zodiacal background-limited space telescope with mirror diameter $D$ and field of view larger than the angular radius $\theta_{\rm CGM}$ would be sensitive to 
\begin{equation}
 I_{\lambda_X} =  30  \,\gamma\, {\rm cm}^{-2}{\rm s}^{-1} {\rm sr}^{-1} \left(\frac{S/N}{10} \right)  \left( \frac{t_{\rm obs}}{100 \, {\rm hr}} \frac{q}{0.3} \frac{\Delta \lambda}{\rm \AA}\right)^{-1/2}  \left(\frac{\theta_{\rm CGM}}{0.5'} \right) \left( \frac{D}{1~{\rm m}} \right)^{-1},
 \label{eqn:sens}
\end{equation}
where $q$ is the detector efficiency, $t_{\rm obs}$ the observing time, and $\Delta \lambda$ the wavelength resolution.\footnote{This assumes the measurement is Zodiacal background-limited with $I_{\rm Zod} = 500  ~\gamma {\rm ~cm}^{-2} ~{\rm s}^{-1} ~{\rm sr}^{-1} {\rm \AA}^{-1}$, which is consistent with the value at high ecliptic latitude reported by \citet{2014Ap&SS.349..165M}.}  UV grism, spectrograph, and IFU designs often reach $\Delta \lambda \sim 1$\,\AA\ or even smaller values, whereas narrowband filters in the UV can have $\Delta \lambda$ as small as tens of Angstrom with existing technology.

Figure~\ref{fig:f1} motivates values for $\alpha_{\lambda_X}$ for a metallicity of $Z= 0.3\, Z_\odot$. We use the {\sf Cloudy} nebular ionization code \citep{cloudy,cloudy2017} (version c17.02) to calculate the fractional intensity of the emission coming out in the most intense lines at a specific temperature, assuming optically thin emission and an extragalactic background field from the fiducial background model in \citet{KS19}.\footnote{A harder ionizing background, calculated by setting the spectral index of quasar specific luminosity to $1.4$ instead of the fiducial $1.8$, only has a minimal effect on the relative emission of the lines (not shown in Fig.~\ref{fig:f1}). The few significant differences seen when implementing this harder ionizing background -- with the largest differences at $n_e=10^{-4.0}$\cmminusthree -- includes: a tripling of the intensity in \OV\ 630\,\AA\ for $10^5$\,K gas, a $\sim 40 \%$ decrease in the intensity of \OV\ 630\,\AA\ for $10^{5.5}$\,K gas, and $\sim 25\%$ increase in the intensity of \HeII\ 304\,\AA\ for $10^{4.5}$\,K gas.} Each panel considers a specific hydrogen density covering a range relevant to that of the CGM. The different marker shapes represent calculations with specific temperatures in the range $10^4 - 10^6$\,K~(see figure legend).  Of course, some of the most intense lines are the Lyman-$\alpha$ transitions of hydrogen and helium (1216\,\AA\ and 304\,\AA\ respectively), with CGM emission in the former likely being dwarfed by scattered light from the host galaxy and the latter being so far in the UV that it must be observed from $z>2$ to avoid Galactic absorption.  Some metal ions can be almost as intense as these primordial species, such as \CIII\ 977\,\AA\ and \CIV\ 1548\,\AA\ for $10^5$\,K gas, and \OV\ 630\,\AA\ and \OVI\ 1032\,\AA\ for $10^{5.5}$\,K gas. Million Kelvin gas emits most prominently in $X$-ray iron lines.  It may come as a surprise not to see \OVII\ and \OVIII\ as a top-five line at $10^6~$\,K -- their helium-like and hydrogen-like structure for which the line excitation energy is comparable to the ionization energy makes them good emitters over a narrow temperature range.  Interestingly, one of the dominant lines at $10^4-10^{4.5}$\,K, especially for denser gas, are the optical \OII\ $3700$\,\AA\ and \OIII\ $5000$\,\AA\ transitions. For metals inhabiting $\gtrsim 10^{4.5}$K gas, a lower density affects the fractional emission by principally increasing each element's ionization.

Let us take the example of \OVI\ 1032\,\AA\ to relate the fractions in Figure~\ref{fig:f1} to rough estimates for the intensities. Let us assume that a fraction $Y$ of the cooling energy from the CGM owes to $\sim 10^{5.5}$\,K gas, where $Y\sim {\cal O}(1)$ would not be unexpected since this is not so different than the virial temperature of Milky Way halos. Figure~\ref{fig:f1} shows that a fraction $\alpha_{\lambda_X} \approx 0.2$ of the emission of $\sim 10^{5.5}$\,K gas comes from \OVI\ 1032\,\AA, with a weak dependence on the density. Taking equation~(\ref{eqn:Ienergy}) and assuming this radiation comes from within $R_{\rm CGM}=100\;$kpc from a galaxy with SFR =${5 M_\odot {\rm ~yr}^{-1}}$ yields $I_{\lambda_X} = 60\, Y ~\gamma {\rm ~cm}^{-2} ~{\rm s}^{-1} ~{\rm sr}^{-1}$,   which equation~(\ref{eqn:sens}) suggests is at the borderline of being detectable with a $D=1\,$m UV space telescope. Furthermore, denser gas in the inner CGM is likely to emit more intensely than this aggregate CGM estimate. 

Figure~\ref{fig:f1} assumes optically thin conditions. This ignores that lines with $\lambda < 912$\,\AA, such as one of our brightest lines \OV\ 630\,\AA, can be reabsorbed by hydrogen continuum absorption if the hydrogen column is above $\sim 10^{17} (\lambda/912{\rm \AA})^{-3} $~cm$^{-2}$.  For the low redshift $<150$~kpc CGM, the hydrogen column exceeds $10^{17}$~cm$^{-2}$ for a third of the sightlines in the COS-Halos sample \citep{prochaska17}, although with decreasing radius or increasing redshift higher columns are anticipated.  Another potentially important source of opacity for $>4\,$Ry background photons is \HeII\ self shielding, which occurs at smaller neutral hydrogen columns of $N_{\rm HI} \approx 1\times10^{16} \times (100/[\Gamma_{\rm HI}/\Gamma_{\rm HeII}])$ cm$^{-2}$  \citep{2009ApJ...694..842M}, where the ratio of the \HI\ and \HeII\ photoionization rates is predicted to be $\Gamma_{\rm HI}/\Gamma_{\rm HeII} \sim 100$ at $z<3$ in ionizing background models \citep{KS19}. Roughly half of HST/COS systems show larger \HI\ columns for which the \HeII\ should self shield to the background.  Such self-shielding results in the \HeII\ recombination emission, which dominates the \HeII\ emission at $\lesssim 10^{4.5}$K in Figure~\ref{fig:f1}, to saturate at intensities below what could be conceivable observed (\S~\ref{sec:resultsHI}).   \HeII\ self shielding also affects ions with ionization potentials near the \HeII\ edge such as \NIII\ and \OIII, resulting in larger columns in these ions at temperatures of $10^4-10^{4.5}$K where these ions are photoionized.  A third possible effect when the optically thin limit does not hold is line self-absorptions.  However, this turns out to not be important:  while empirically low-redshift CGM lines have optical depths of $\lesssim 1$ (although the resolution of HST/COS is an issue with this determination), even for cases when the lines become optically thick, absorption results in re-emissions in the same transition for all the important ground state metal transitions we consider, with the exception of the \HI\ Lyman-series (\S~\ref{sec:resultsHI}).

Given these preliminary estimates, we now consider in more detail the atomic physics of CGM emission and relate the predicted intensity to column density measurements from absorption-line observations.

\begin{table}[h]
\label{Table1}
\def\arraystretch{1.3}
\centering
        \begin{tabular}{c c c|c |c |c}
        \hline
  \textbf{Species} & $\boldsymbol{\lambda}$ [\AA] & \textbf{Transition} & ${\boldsymbol{\cal B}}_{\boldsymbol{\lambda_X}}$ & $\mathbf{[\Upsilon/g_{\rm grd}]}$
  & \textbf{Notes}\\
            \hline
        \multirow{2}{*}{\CIII} & \multicolumn{1}{l}{978} & $  ^{1}P_{1}$ \(\rightarrow\) $ ^{1}S_{0}$ & 1.0 & [3.19, 3.42, 3.99, 5.27, 7.54] &likely strongest line of $10^5$K gas  \\\cline{2-6}
                                & \multicolumn{1}{l}{1907$^{\lambda\lambda}$} & 
                                 $^{3}P_{2}$ \(\rightarrow\) 
                                 $ ^{1}S_{0}$ & 1.0 & [0.41, 0.47, 0.44, 0.31, 0.18] & $\lambda\lambda$(1909\,\AA; 0.60) \\\cline{1-6}
        \multirow{1}{*}{\CIV} & \multicolumn{1}{l}{1548$^{\lambda\lambda}$} & $  ^{2}P_{3/2}$ \(\rightarrow\) $ ^{2}S_{1/2}$ & 1.0 & [2.86, 2.97, 3.25, 3.82, 4.83] & $\lambda\lambda$(1551\,\AA; 0.52); from $10^5$K, relatively bright  \\\cline{1-6}
        \multirow{2}{*}{\NII} & \multicolumn{1}{l}{1084} &  $^{3}D_{1}$ \(\rightarrow\) $ ^{3}P_{0}$ & 0.56 & [0.39, 0.50, 0.69, 0.91, 1.26] &-- \\\cline{2-6}
                                 & \multicolumn{1}{l}{6583$^{*}$} & 
                                 $^{1}D_{2}$ \(\rightarrow\) 
                                 $ ^{3}P_{2}$ & 0.75 & [0.28, 0.29, 0.29, 0.30, 0.30] &  \(\Upsilon\)(6527\,\AA); few$\times$weaker than 5007\,\AA\ at fixed $nT$  \\\cline{1-6}
        \multirow{1}{*}{\NIII} & \multicolumn{1}{l}{990} & $  ^{2}D_{3/2}$ \(\rightarrow\) $ ^{2}P_{1/2}$ & 0.84 & [0.89, 0.97, 1.08, 1.27, 1.68] & -- \\\cline{1-6}
        \multirow{1}{*}{\NIV} & \multicolumn{1}{l}{765} & $  ^{1}P_{1}$ \(\rightarrow\) $ ^{1}S_{0}$ & 1.0 & [3.20, 3.35, 3.60, 4.20, 5.50] & --   \\\cline{1-6}
        \multirow{1}{*}{\NV} & \multicolumn{1}{l}{1239$^{\lambda\lambda}$} & $  ^{2}P_{3/2}$ \(\rightarrow\) $ ^{2}S_{1/2}$ & 1.0 & [2.16, 2.21, 2.34, 2.66, 3.21]   &  $\lambda\lambda$(1242\,\AA; 0.50) \\\cline{1-6}
        \multirow{1}{*}{\OII} & \multicolumn{1}{l}{3726$^{\lambda\lambda}$} & $  ^{2}D_{3/2}$ \(\rightarrow\) $ ^{4}S_{3/2}$ & 1.0 & [0.14, 0.14, 0.15, 0.15, 0.16]   &  $\lambda\lambda$(3729\,\AA; 1.48); probes denser gas $\gtrsim 10^{-2}$\cmminusthree \\\cline{1-6}

            \multirow{3}{*}{\OIII} & \multicolumn{1}{l}{2321} &  $^{1}S_{0}$ \(\rightarrow\) $ ^{3}P_{1}$ & 0.13 & [0.03, 0.04, 0.04, 0.03, 0.02] & -- \\\cline{2-6}
                                 & \multicolumn{1}{l}{4959$^{*}$} &  $^{1}D_{2}$ \(\rightarrow\) $ ^{3}P_{1}$ & 0.25 & [0.25, 0.29, 0.29, 0.23, 0.15] &  $\Upsilon$(4931\,\AA) \\\cline{2-6}
                                 & \multicolumn{1}{l}{5007$^{*}$} &  $^{1}D_{2}$ \(\rightarrow\) $ ^{3}P_{2}$ & 0.75 & [0.25, 0.29, 0.29, 0.23, 0.15]  &  $\Upsilon$(4931\,\AA); {\footnotesize strongest optical line; probes $10^{4.3}$K}  \\\cline{1-6}
        \multirow{1}{*}{\OIV} & \multicolumn{1}{l}{555} &   $^{2}P_{1/2}$ \(\rightarrow\) $^{2}P_{3/2}$ & 0.34 & [0.24, 0.23, 0.24, 0.27, 0.34]& --   \\\cline{1-6}
        \multirow{1}{*}{\OV} & \multicolumn{1}{l}{630} &   $^{1}P_{1}$ \(\rightarrow\) $^{1}S_{0}$ & 1.0 & [2.32, 2.47, 2.65, 3.01, 3.75]  & brightest line of $10^{5}-10^{5.5}$K gas   \\\cline{1-6}
        \multirow{1}{*}{\OVI} & \multicolumn{1}{l}{1032$^{\lambda\lambda}$} &   $^{2}P_{3/2}$ \(\rightarrow\) $^{2}S_{1/2}$ & 1.0 & [1.65, 1.68, 1.76, 1.95, 2.30] & $\lambda\lambda$(1038\,\AA; 0.50); $I_{\text{CE}}$ weak $T$-dep. for $\gtrsim 10^5$K\\\cline{1-6}
        \multirow{1}{*}{\SiIII} & \multicolumn{1}{l}{1207} & $  ^{1}P_{1}$ \(\rightarrow\) $ ^{1}S_{0}$ & 1.0 & [5.81,  7.00, 8.76, 12.6, 19.1] & -- \\\cline{1-6}
        \multirow{1}{*}{\SiIV} & \multicolumn{1}{l}{1394$^{\lambda\lambda}$} & $  ^{2}P_{3/2}$ \(\rightarrow\) $ ^{2}S_{1/2}$ & 1.0 & [5.20, 5.30, 5.75, 7.05, 9.50]  & $\lambda\lambda$(1403\,\AA; 0.50)  \\\cline{1-6}
        \end{tabular}
\caption{Atomic information, branching ratios, effective collision strengths, and major conclusions for various significant emission lines.  The collision strengths can be used to easily calculate line emission properties via equations~(\ref{eqn:collem}) and (\ref{eqn:inten&emis}).  The effective collision strengths, $\Upsilon/g_{\rm grd}$, are from the ion's ground state to the excited state that leads this transition.  Each entry in the $\Upsilon/g_{\rm grd}$ vector is evaluated at temperatures of [${10}^{4}$, $10^{4.5}$, ${10}^{5}$, $10^{5.5}$, ${10}^{6}$]$~$K. ${\cal B}_{\lambda_X}$ is the fraction of spontaneous transitions from the upper state that result in the line.
Transitions that do not end up in the ground fine structure state are indicated with $*$ in the wavelength column; the excitation rate $\Upsilon$ that is listed for these transitions is instead a different transition whose wavelength is indicated in the notes column. While doublet lines are denoted by ${\lambda\lambda}$ and that line's mate is found in the notes column. Following the corresponding doublet line in the notes is a conversion factor to multiply the $\Upsilon/g_{\rm grd}$ to get the corresponding doublet line's $\Upsilon/g_{\rm grd}$. For more details see Appendix~\ref{ap:linestrengths}. }\label{table1}

\end{table}

\subsection{Emission Line Physics}
\label{sec:line_emission}

In this section we show that when low-redshift CGM line emission is at an observable level, the emission process \emph{must} be from collisions unless the line is from \HI.  The only exception is if the radiation field in the proximity of the galaxy is significantly stronger than the extragalactic background assumed in our calculations (see \citealp{2018ApJ...869..159U}).

In a homogeneous slab approximation, the \emph{CGM-frame} intensity from collisionally excited emissions in the transition $\lambda_X$ is easily computed from the column density of the ion in question $N_X$ and the density $n_e$ and temperature $T$ of the exciting electrons. Namely,
\begin{equation}
    I_{\rm CE} = \frac{1}{4\pi} {\cal B}_{\lambda_X} \langle \sigma_{\lambda_X} v \rangle_T  n_e N_{X} = 80 {~\gamma\, {\rm  cm}^{-2} {\rm s}^{-1} {\rm sr}^{-1}~}  {\cal B}_{\lambda_X} \frac{\langle \sigma_{\lambda_X} v  \rangle_T }{10^{-7}\;{\rm cm^{3}\;s}}  \left( \frac{n_e}{10^{-4}\; {\rm cm}^{-3}} \right) \left( \frac{N_X}{10^{14} \; {\rm cm}^{-2}} \right),
    \label{eqn:collem}
\end{equation}
where ${\cal B}_{\lambda_X}$ is the fraction of spontaneous transitions from the excited state of the line of interest, which is always a substantial fraction of unity, if not unity, for the transitions we consider.\footnote{Equation~(\ref{eqn:collem}) ignores collisional excitations to a higher energy level than the $\lambda_X$ transition. This approximation is good for the metal ions we study, for which other electronic orbital configurations are at much higher energies, as confirmed by the comparison of our estimates with {\sf Cloudy} calculations (Appendix~\ref{ap:linestrengths}). However, it can be less accurate for other hydrogen and helium as considered in \S~\ref{sec:resultsHI}.} For specific ions we will generalize this expression beyond the homogeneous slab approximation. The physics is buried in $\langle \sigma_{\lambda_X} v  \rangle_T$, the thermally-averaged collisional excitation cross section times the electron velocity.  A simple estimate for this quantity is
\begin{eqnarray}
    \langle \sigma_{\lambda_X} v \rangle_T &=& \sqrt{4\pi} a_0^2 \, \alpha c \sqrt{\frac{Ry}{kT}}~ [\Upsilon/{g_{\rm grd}}] ~\exp{\left[\frac{-E_{X}}{k T}\right]};  \\ &=& 2.7\times10^{-8}~ [\Upsilon/{g_{\rm grd}}] ~\left(\frac{T}{10^5 {\rm K}}\right)^{-1/2}  \exp{\left[\frac{-E_{X}}{k T}\right]} {\rm ~~cm^{3}\; s^{-1}},
\end{eqnarray}
where $E_X = h c/\lambda_X$ is the transition energy, $a_0$ is the Bohr radius, $Ry$ the Rydberg energy, $\alpha$ is the fine structure constant, and $[\Upsilon/{g_{\rm grd}}]$ a fudge factor needed to fit the actual coefficient \citep[e.g.][]{2006agna.book.....O}. This generally order-unity factor is listed in Table~\ref{table1} for the transitions we consider in this work. See Appendix~\ref{ap:linestrengths} for more details. Thus, once the gas temperature satisfies $k \,T>E_{\rm X}$, or $k \, T \sim 10^5$\;K for UV transitions, $\langle \sigma_{\lambda_X} v \rangle_T$ is relatively constant with $T$, and it falls quickly to zero as $T$ decreases below this threshold.

Any recombination emission (driven by collisional- or photo-ionization) from metals is likely to be undetectably small.  An estimate for the ratio of the intensity from radiative recombinations to collisional ionizations is:
\begin{equation}
\frac{I_{\rm REC}}{I_{\rm CE}} = \frac{f_{\lambda_X} \langle \sigma_{\rm rec} v \rangle_T}{\langle \sigma_{\lambda_X} v\rangle_T } \frac{N_{XI}}{N_X} \sim \left(\frac{f_{\lambda_X}}{Z_{\rm sc}\,[\Upsilon/{g_{\rm grd}}]}\right)   \left(\frac{E_{{\rm ion}, X}}{m_e c^2} \right)^{3/2}  \frac{N_{XI}}{N_X} \exp \left(\frac{E_{X}}{kT}\right) ,
\label{eqn:IrecICe}
\end{equation}
where $N_{XI}$ is the column in the next ionization stage above $X$, $E_{{\rm ion}, X}$ the ionization potential of ion $X$, and $f_{\lambda_X}$ is the fraction of $XI + e \rightarrow X$ recombinations that produce a photon in transition ${\lambda_X}$, and the rightmost relation uses that $\langle \sigma_{\rm rec} v \rangle_T \sim  \alpha c (a_0/Z_{\rm sc})^{2} \times (E_{{\rm ion}, X}/[m_e c^2])^{3/2} \times  (E_{{\rm ion}, X}/kT)^{1/2}$ for $k T\ll E_{{\rm ion}, X}$, with $Z_{\rm sc}$ the screened nuclear charge, a rough approximation motivated by direct to ground recombination for hydrogenic atoms \citep{rybicki}. As ${f_{\lambda_X}}/(Z_{\rm sc} \,[\Upsilon/{g_{\rm grd}}]) \sim 1$ for the most promising lines and ${E_{{\rm ion}, X}}/{m_e c^2} \sim 10^{-5}$, equation~(\ref{eqn:IrecICe}) suggests that the ratio $ I_{\rm REC}/{I_{\rm CE}}$ is much less than one unless the gas temperature is much smaller than $E_X/k$ or the fraction of the element in ion $X$ is small.  In particular, satisfying ${I_{\rm REC}}/{I_{\rm CE}} \gtrsim 1$ at high temperatures where $E_X \lesssim kT$ requires that ${N_{XI}}/{N_X} \gtrsim 10^6$ -- a death sentence for any metal ion $X$ to be detectable as it will be in other ionization stages than $X$ and $XI$.  If instead ${N_{XI}}/{N_X} \sim 1$, as necessary for substantial emission in $X$, the temperatures need to be low with $E_X/kT \gtrsim 10$.  The latter requirement for the UV transitions we investigate, which have $E_X \sim 10^5$K/$k$, means recombination emission is the dominant process at temperatures near the minimum that is achievable for atomic gas, $T\sim 10^4$K.\footnote{For optical transitions with energies of a few eV, collisional ionizations are always likely to be dominant.}  Recombination emission at $T\sim 10^4$K is driven by photoionizations such that each recombination is in balance with a photoionization. This fact provides a second way to estimate the \emph{CGM-frame} recombination emission: 
\begin{eqnarray}
I_{\rm REC} &\approx& \frac{f_{n'n} \Gamma_X(z)}{4\pi \sigma_{\rm eff}}\;\left(1- \exp\{-\sigma_{\rm eff} N_{X}\} \right), \label{eqn:photoX} 
\end{eqnarray}
 where $\sigma_{\rm eff} $ is the effective photoionization cross section and $\Gamma_X(z)$ is the photoionization rate for ion $X$.\footnote{While the value of $\sigma_{\rm eff}$ only matters in the optically thick limit $N_{X} \gtrsim \sigma_{\rm eff}^{-1} $, which likely does not apply to any metal and often not even hydrogen, we choose the effective cross section from averaging the cross section over the spectrum of ionizing radiation (\S~\ref{sec:resultsHI}).}  This choice, which is only relevant in \S~\ref{sec:resultsHI}, negligibly affects our results. \label{foot:sigeff}
 The background photoionization-balancing emission is likely only detectable for hydrogen lines because hydrogen absorbs the bulk of the ionizing photons. As the Ly$\alpha$ forest constrains $\Gamma_{\rm HI, bkgd}$ to be $ \approx 4\times10^{-14} (1+z)^5$\,s$^{-1}$ at $z\lesssim 3$ \citep{2015ApJ...811....3S, KS19}, this results in a maximum of $I_{\rm REC} \approx 10^3 [(1+z)/1.3]^5~\fphosr$ for hydrogen lines, and an unobservably small intensity that is at least hundreds of times smaller for helium and metal lines (as their ions absorb far fewer continuum ionizing photons). We solidify this argument in \S~\ref{sec:resultsHI}.

\section{empirical emission models}
\label{sec:results}

We now use CGM observations to estimate emission line intensities and compare them to the detection limits of existing optical instruments and potential UV space telescopes. We consider emission in some of the most promising lines discussed in \S~\ref{sec:physics} (see Table~\ref{Table1}): (i) \OVI~1032\,\AA, a useful line for its active emission from the warm/hot gas it likely inhabits (\S~\ref{ss:OVI}); (ii) optical lines, which are the most promising collisional emission from  cool gas (\S~\ref{ss:optical}); (iii) other UV lines sourced by gas at intermediate temperatures of $\sim 10^5$~K (\S~\ref{ss:BL}); and (iv) fluorescent emission off the extragalactic UV background, only detectable from hydrogen lines (\S~\ref{ss:HI}). 

\subsection {\OVI\ Emission}
\label{ss:OVI}
Let us start our census with a pair of the most promising lines, the \OVI\ 1032, 1038\,\AA\ doublet. Of the lines we discuss, this doublet is perhaps the simplest to relate to the CGM gas since $N_{\rm OVI}$ likely exists at temperatures where $\langle \sigma v \rangle_T$ is relatively constant with a value of $\approx 2\times10^{-8}$~cm$^{3}$~s$^{-1}$ for the $1032$\,\AA\ line and half this value for the $1038$\,\AA\ line (see Table~\ref{Table1} and Figure~\ref{fig:fandsigmav} in the Appendix). This contrasts with other UV transitions arising from ions that trace lower density gas. The \OVI\ 1032\,\AA\ line's nearly temperature-independent $\langle \sigma_{\lambda_X} v \rangle_T$  allows us to generalize equation~(\ref{eqn:collem}) to the case of a sightline with varying temperature, density and metalicity:
\begin{equation}
    I_{1032A} \approx 50~\fphosr~ \left( \frac{\langle  n_e \rangle_{\rm OVI} }{10^{-4}\; {\rm cm}^{-3}} \right) \left( \frac{N_{\rm OVI}}{3\times 10^{14} \; {\rm cm}^{-2}} \right) (1+z)^{-3},
    \label{eqn:OVI}
\end{equation}
where $\langle  n_e \rangle_{\rm OVI}$ is the \OVI\ density-weighted average, namely $\langle  n_e \rangle_{\rm OVI} = \int dr \, n_e n_{\rm OVI}/\int dr \,n_{\rm OVI}$ where the integral is over the sightline across a halo.  This simplifies from the exact expression $\langle  n_e \rangle_{\rm OVI} = \int dr  \,\langle  \sigma v \rangle_{T(r)} n_e(r) n_{\rm OVI}(r)/[\langle  \sigma v \rangle_{\rm ref}  N_{\rm OVI}]$, where $\langle  \sigma v \rangle_{\rm ref}$ is the reference cross section that will cancel out when evaluating the intensity and $N_{\rm OVI} = \int dr \, n_{\rm OVI}(r)$, because of the near temperature-independence of $\langle  \sigma v \rangle_T$ over relevant temperatures.  Since we are tying our measurements to observed columns, only $\langle  n_e \rangle_{\rm OVI}$ is required as a further input to estimate the intensity.

\begin{figure}
    \centering
    \includegraphics[width=18cm]{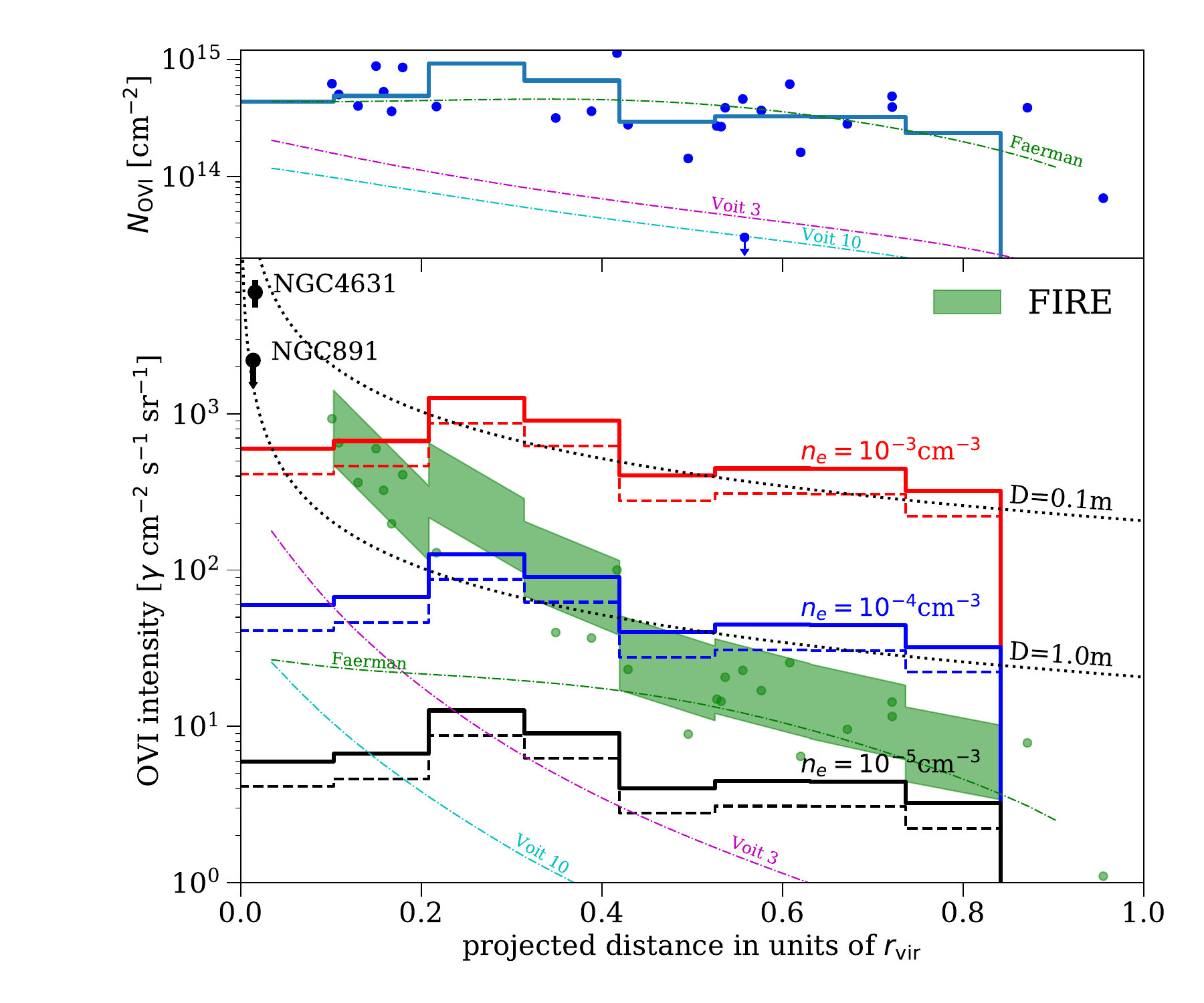}
    \vspace{-.3cm}
    \caption{ {\bf Top panel:} HST/COS measurements of the \OVI\ column density for sightlines through $\approx 10^{12}M_\odot$ halos. The histogram is the logarithmic average of the COS-Halos sample \citep{werk14}, and the blue points are the re-analysis of \citet{johnson17} that selected isolated star-forming $\sim L_\star$ galaxies out to larger radii. The thin curves show the \OVI\ column in the theoretical models of \citet{voit19}, for cooling to dynamical time ratios of 3 and 10, and also of \citet{faerman20}. 
    {\bf Bottom panel:} Empirical estimates for the CGM-frame surface brightness in the \OVI\ 1032\,\AA\ line. We first show constant density and temperature models.  The solid histograms assume $10^6$K gas and the dashed $10^5$K, with both densities at the specified densities and using the histogram for $\langle \log N_{\rm OVI} \rangle$ shown in the top panel (see eqn.~\ref{eqn:OVI}).  The temperatures are chosen just to illustrate the near temperature-independence of the emission.  The green highlighted region shows a model that again uses the histogram in the top panel for $\langle \log N_{\rm OVI} \rangle$ and takes the average density profile taken from the FIRE simulation (\citealp{2019MNRAS.488.1248H}). The shaded band allows for a factor of three higher than the model's virial density, motivated by \OVI\ tracing gas with temperatures up to a factor of three smaller than the virial temperature. The green markers use the $N_{\rm OVI}$ measurements shown in the top panel rather than the average and for a density that is 1.5 times the models' virialized gas density. This model assumes that the \OVI\ ionization fraction is constant independent of density and radius, as occurs in collisional ionization equilibrium and for a single temperature. The thin curves in both panels are the more sophisticated models of \citet{voit19} and \citet{faerman20}, where the temperature is a function of radius.  These models produce the corresponding column in the top panel and are not noramlized to COS-Halos. The black points with error bars are the FUSE measurements of \OVI\ in emission just above the disk of two spiral galaxies NGC4631 and NGC891 \citep[upper and lower points respectively]{2021arXiv210305008C}. The black dotted curves are the $10\sigma$ sensitivities for a $100$hr observation with $D=10\,$cm and $D=100\,$cm telescopes with $1\,$\AA\ spectral resolution, assuming the noise is set by Zodiacal foregrounds, 30\% detector efficiency, and systems at $z=0.2$ (see details in \ref{ss:OVI}).}
    \label{fig:OVI}
\end{figure}

Figure~\ref{fig:OVI} shows more detailed estimates for the surface brightness of this line based on the HST/COS column density measurements. The top panel shows HST/COS constraints on \OVI\ column density for $\approx 10^{12}~M_\odot$ halos to the virial radius, $r_{\rm vir}$. The histogram is the logarithmic average of the COS-Halos measurements \citep{werk14}, and the blue points are the re-analysis of the data from \citet{johnson17} for isolated star-forming $\sim L_\star$ galaxies, extending to larger projected radii. The bottom panel shows empirical estimates for the predicted emission-frame surface brightness in the \OVI\ 1032\,\AA\ line computed using this histogram for $\langle \log N_{\rm OVI} \rangle$, and assuming different gas density and temperature profiles.

For intuition, first consider constant density models with $n_e = 10^{-3}$, $10^{-4}$, and $10^{-5}$\cmminusthree. The dashed curves assume $10^5$~K gas while the solid histograms assume $10^6$~K gas, a temperature range that is broader than that inhabited by the \OVI\ in all collisionally ionized models.\footnote{We acknowledge that these are not the most physical temperatures for collisionally ionized \OVI. The differences are even smaller if we choose $10^{5.5}$K and $10^{6.2}$K, the two characteristic temperatures for collisionally ionized \OVI.} The similarity of these predicted intensities at these two temperatures shows that the emission in collisionally ionized models is insensitive to temperature and primarily sensitive to density. At lower temperatures, that could occur if the \OVI\ is photoionized (requiring extremely low densities), the emission is exponentially suppressed; \OVI\ in $T\ll10^5$~K gas emits negligibly. A density of $n_e = 10^{-4}$\cmminusthree is approximately the average density required for a $z\approx 0.2$ halo to contain all baryons within the virial radius, but if the \OVI\ is associated with $10^{5.5}$~K gas, as in most models, pressure confinement with the hotter virialized gas could enhance the density.
 
The black dotted curves show the $S/N=10$ Zodiacal background--limited sensitivities when averaging the intensity over a circle that spans the projected distance, for two UV telescope diameters, $D$, with $1$\,\AA\ spectral resolution and 30\% detector efficiency, and exposure time of $100$~hours (c.f. eqn.~\ref{eqn:sens}). A cubesat-sized telescope with $D=10$\,cm is sensitive to \OVI\ if the observed column inhabits gas with relatively high densities of $n_e = 10^{-3}$\cmminusthree, as may occur towards the inner CGM. A NASA explorer-class--like satellite with $D= 100\,$cm is sensitive to the intensity of \OVI\ that inhabits densities as low as $n_e = 10^{-4}$\cmminusthree.

The green shaded region shows a more physically motivated model for the radial density profile of virialized gas that, as with the single-density models, is constrained to have the HST/COS \OVI\ columns shown in the top panel. This model takes as input the $z=0.2$ gas density profile for a $10^{12}~M_\odot$ halo in the FIRE zoom-in galaxy formation simulation, which is well fit by $n_{e}^{\rm vir}(r) = 1.5\times10^{-4}(r/0.3 r_{\rm vir})^{-2}$\cmminusthree \citep{2019MNRAS.488.1248H}. Our calculation assumes the \OVI\ ionization fraction is constant independent of density and radius, as occurs in collisional ionization equilibrium and for a single temperature, and we integrate in the line-of-sight direction out to infinity to calculate  $\langle  n_e \rangle_{\rm OVI}$ and hence the projected emission, using additionally the $N_{\rm OVI}$ histogram shown in the top panel and equation~(\ref{eqn:OVI}). The width of the shaded band allows for a factor of three higher density over $n_{e}^{\rm vir}$ -- motivated by \OVI\ tracing gas with temperatures a factor of three smaller than the $10^6$\,K virial temperature. The resulting emission is above the threshold for a $D=100$~cm telescope out to $0.4~r_{\rm vir}$, or $100$~kpc. We have also looked at a second model based on empirical constraints for the Milky Way halo gas (see \citealp{2015ApJ...800...14M,voit19}) and find emission that is a factor of two smaller than we find for the FIRE density-profile model.
 
The green points in the bottom panel of Figure~\ref{fig:OVI} show estimates based on the individual HST/COS $N_{\rm OVI}$ measurements shown in the top panel, where to map them to an intensity $\langle  n_e \rangle_{\rm OVI}$ is taken to be $1.5\times$ the FIRE line-of-sight averaged virialized density. The scatter in these values may indicate how much halo-to-halo variance is present, but also could owe to structures within individual halos.

The assumption that the \OVI\ traces all gas proportional to density may not hold as the \OVI\ abundance is very temperature sensitive (unlike its emission at fixed $n_{\rm OVI}$). If a radially varying temperature results in \OVI\ tracing gas that is at the outskirts of halos, then we expect the emission to be lower than the green shaded region. For example, in the \citet{faerman20} model, the gas temperature is lower at large radii, resulting in most of the \OVI\ residing at $\gtrsim 150$\,kpc. The thin green curve in Figure~\ref{fig:OVI} shows the emission in this model, which has intensities of $10-30{~\gamma\, {\rm cm}^{-2}~{\rm s}^{-1}~{\rm sr}^{-1}}$. Next, the thin cyan and magenta curves show the model of \citet{voit19}, which also has a temperature gradient, for a uniform $0.5Z_\odot$ and cooling to dynamical time ratios of $t_{\rm cool}/t_{\rm dyn}=3$ and 10, respectively. In the \citet{voit19} model, the gas density profile is completely specified by these parameters plus the NFW potential, as this ratio is required to hold at all radii.  These models undershoot the OVI column density measurements at large radii, contributing to their fast decrease in the line intensity with increasing radius.\footnote{\citet{voit19} invokes a broad lognormal distribution of temperatures to correct this undershoot, and we anticipate that if the amount of $\sim 10^{5.5}$\,K gas in this distribution does not vary substantially with radius, the predictions of such a model that matches the observed column density profile would be closer to the FIRE model.}

Finally, the black points with error bars are the FUSE measurements of \OVI\ in emission just above the disk of two spiral galaxies NGC4631 and NGC891 \citep[upper and lower points respectively]{2003ApJ...591..821O, 2021arXiv210305008C}.  The detection in NGC4631 implies $\left({\langle  n_e \rangle_{\rm OVI} }/{10^{-2}\; {\rm cm}^{-3}} \right) \left( {N_{\rm OVI}}/{10^{15} \; {\rm cm}^{-2}} \right) = 0.3$ at $10~$kpc, and if the \OVI\ column at this impact parameter through NGC4631 is similar to the COS-Halos columns of $\sim 10^{15}$\,cm$^{-2}$, this observation is consistent to a factor of $\sim 3$ with the extrapolation of the $r^{-2}$ FIRE density profile to these small radii. Perhaps an even more interesting case is the starburst galaxy  SDSS~J115630.63+500822.1, where \OVI\ emission at $10\,$kpc is detected with $I_{1032A}\approx 50,000~\fphosr$, and a column of $N_{\rm OVI}=10^{15}$\,cm$^{-2}$ is estimated after doubling the column inferred from self-absorption of the galaxy continuum. With these inputs, equation~(\ref{eqn:OVI}) suggests $n_e\approx0.03~$cm$^{-3}$, a factor of few higher than the FIRE extrapolation. This corrects the inference of \citet{hayes}, which inferred $20\times$ larger densities. We further find the absorption arises over a distance $d_{\rm OVI} = 130 (f_{\rm OVI}/0.1)^{-1} (Z/Z_\odot)^{-1}~$pc, a size that makes less certain their conclusion that \OVI\ is driven by boundary layers. We also note that our equation~(\ref{eqn:OVI}) simplifies the \citet{hayes} analysis, which involved ionization modeling and the assumption that the gas is at $10^{5.5}$~K (but returns identical results to ours if done correctly).  Fixing the \OVI\ column, \OVI\ emission from $T>10^5$~K~gas is not sensitive to these assumptions.

\subsection{Optical Line Emission} 
\label{ss:optical}
The lower transition energy of optical lines potentially means that they can be excited collisionally in $\sim 10^4$\,K photoionized gas that is thought to harbor the HST/COS intermediate ionization absorbers. The optical lines of most interest are well known from from studies of ISM emission \citep{1981PASP...93....5B}, namely \OIII\ 5007\,\AA\ and \NII\ 6583\,\AA, although we caution that the physics which shapes them is much different in the CGM. The \OIII\ 5007\,\AA\ line is particularly interesting because of the large columns of \NIII\ measured with HST/COS. These suggest that \OIII\ -- which does not have NUV or FUV allowed ground state radiative transitions and therefore cannot be observed in absorption  -- should have $N_{\rm OIII} \sim 10^{15}-10^{16}$\,cm$^{-2}$. Since for the $5007$\,\AA\ line $\langle \sigma v \rangle_T \approx 3\times10^{-9} \times (T/15,000~{\rm K})$\,cm$^{3}$~s$^{-1}$ for the relevant range of $(1.2-4)\times 10^4$\,K to $\approx 30$ accuracy, equation~(\ref{eqn:collem}) generalizes to
\begin{equation}
    I_{5007A} \approx 160~\fphosr~\left( \frac{\langle  n_e T \rangle_{\rm OIII} }{10^{-3}\; {\rm cm}^{-3} \times 15,000 {\;\rm K}} \right)  \left( \frac{N_{\rm OIII}}{10^{15} \; {\rm cm}^{-2}} \right) (1+z)^{-3}.
    \label{eqn:OIII}
\end{equation}
See the discussion after equation~(\ref{eqn:OVI}) for a close analogy concerning how this equation is derived. This formula overshoots by only a factor of two by $70,000$K, close to the hottest temperatures where \OIII\ can exist, and by a similar factor at $8000$\,K.\footnote{~The equilibrium temperature is $12,000$\,K for $10^{-3}$\cmminusthree and $0.3~Z_\odot$ gas at $z=0.2$, roughly the median specifications of the COS-halos sample \citep{prochaska17}.} Similarly for \NII~6583\,\AA\ the intensity is
\begin{equation}
    I_{6583A} \approx 90~\fphosr~ \left( \frac{\langle  n_e T \rangle_{\rm NII} }{10^{-3}\; {\rm cm}^{-3} \times 15,000 {\;\rm K}} \right)  \left( \frac{N_{\rm NII}}{10^{14.5} \; {\rm cm}^{-2}} \right) (1+z)^{-3},
    \label{eqn:NII}
\end{equation}
valid over $8000-30,000$\;K and within a factor of two at $50,000$\;K. Because of the $n_e T$ scaling over the range in temperature at which these ions exist in the CGM, the emission from both transitions at fixed ionic column density measures pressure. Even though these formulas are collisional emission, they make no assumption about what sets the ionization of the \NII\ and \OIII. For the COS-Halos CGM systems we will focus on, the ionization is set by photoionizations and not collisions, but this optical emission owes to collisional excitation of their low lying states.

\begin{figure}
    \centering
    \includegraphics[height=12cm]{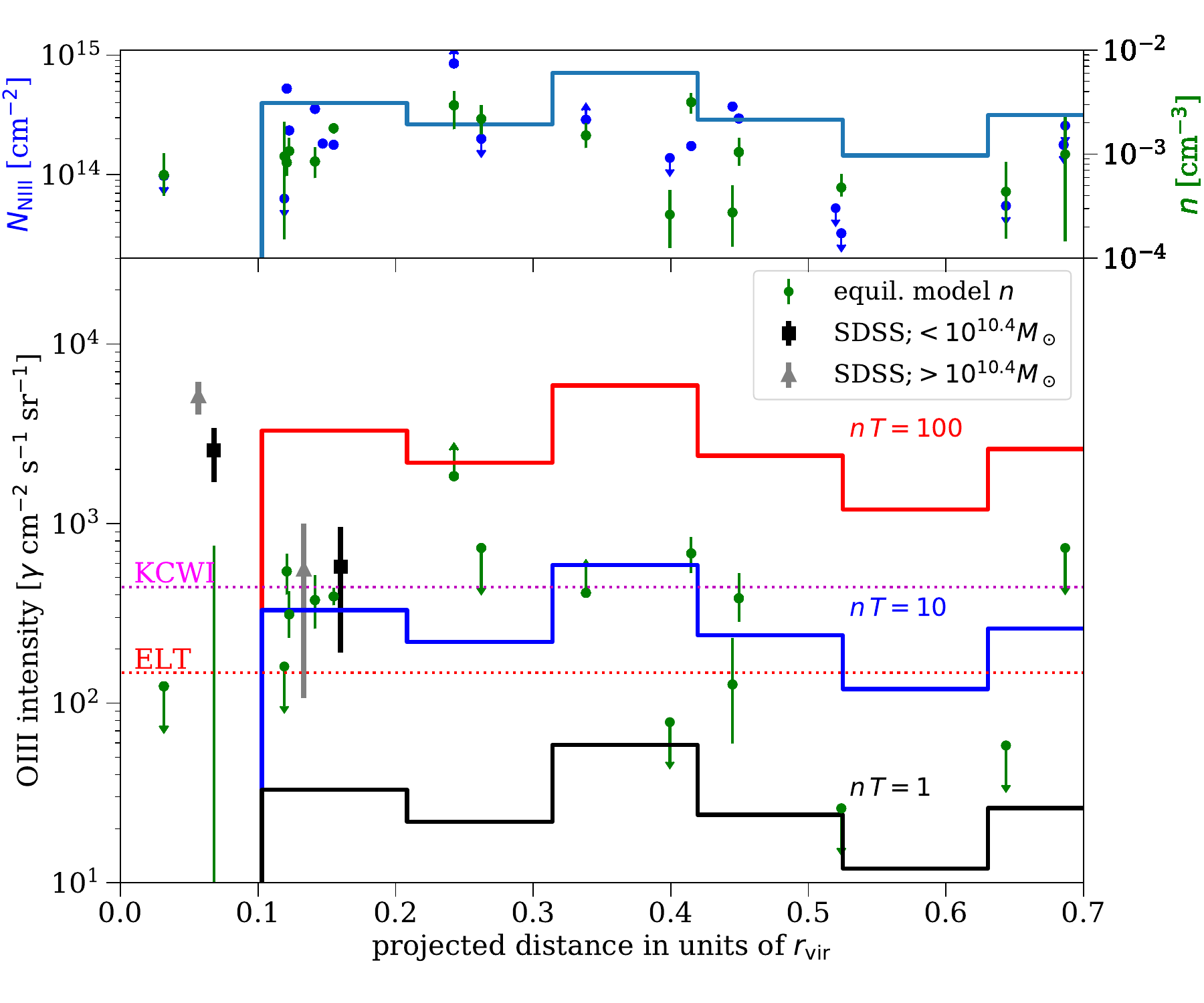}
        \vspace{-.3cm}
    \caption{{\bf Top panels:} HST/COS constraints on \NIII\ column densities for $\approx 10^{12}M_\odot$ halos \citep{werk14} (blue points and histogram) as well as the COS-inferred density from slab equilibrium ionization modeling (green points).  {\bf Bottom panels:}  Empirical estimates for the emission-frame surface brightness in \OIII\ 5007\,\AA. This estimate is based on the \OIII\ column shown in the top panel and that the fraction of nitrogen and oxygen that is twice ionized is the same (as the ionization fraction should be very similar as discussed in the text).  The solid histograms are calculated using the histogram in the top panel for the \OIII\ column and assuming different thermal pressures as \OIII\ emission probes pressure for a fixed column.  The green points do the same estimate but use the density inferences shown in the top panel.  The grey (black) errorbars are the measurements of \citet{2018ApJ...866L...4Z} from stacking SDSS galaxies with $M_* < 10^{10.4} M_\odot$ ($M_* > 10^{10.4} M_\odot$), using an ``average'' virial radius of 250~kpc (300~kpc) to choose the $x$-axis value.  The magenta horizontal dotted line is the Keck/KCWI $10\sigma$ sensitivity assuming the instrument specification of the $0.01$\% sky removal over 600 sq.~arcmin and redshifted to the $z=0.2$ source frame \citep{2010SPIE.7735E..0MM}. The lower horizontal line assumes a factor of three improvement in sensitivity as may be appropriate for an IFU on an ELT.}
    \label{fig:OIII}
\end{figure}

\begin{figure}
    \centering
        \includegraphics[height=12cm]{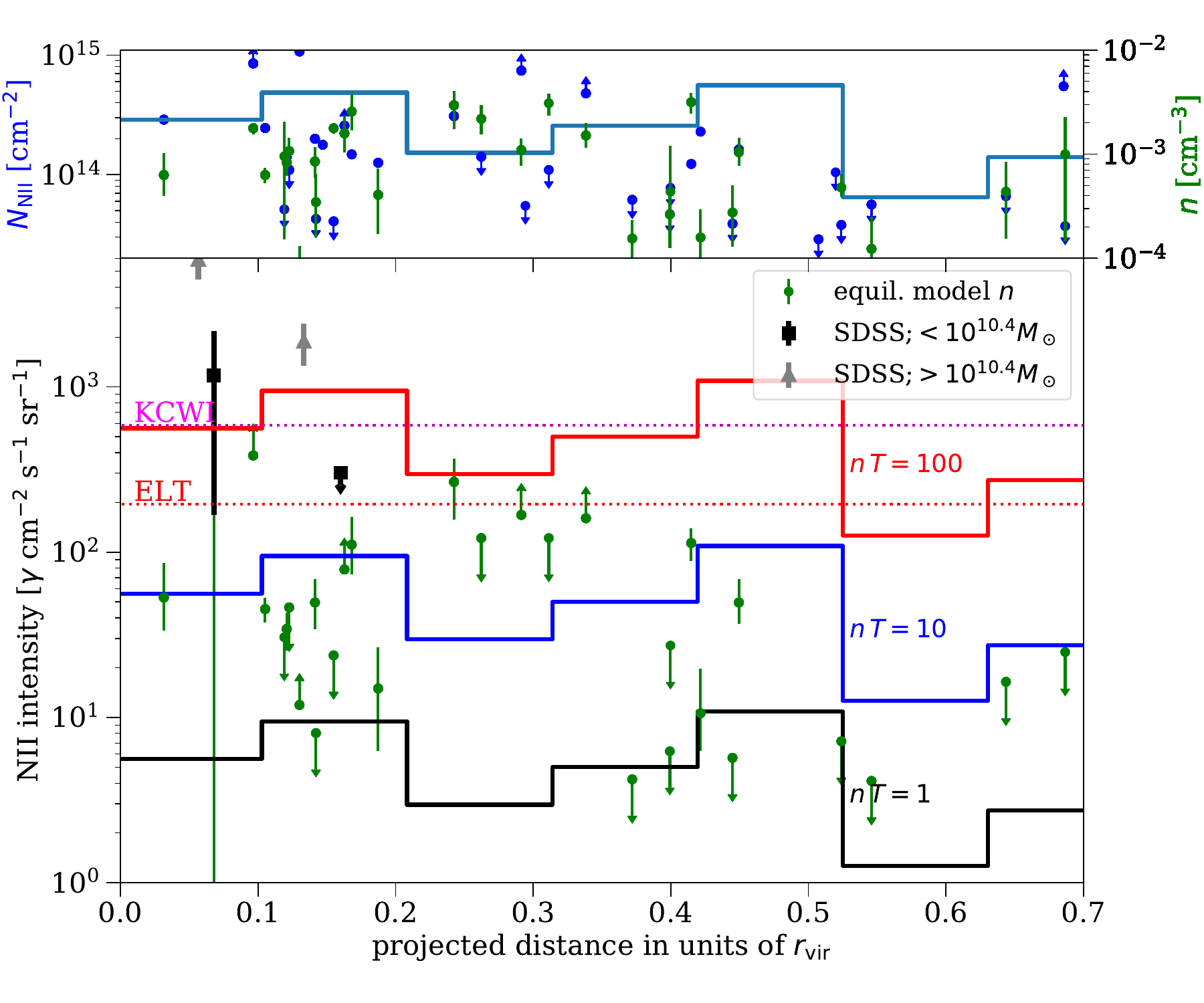}
        \vspace{-.3cm}
    \caption{{\bf Top panels:} HST/COS constraints on \NII\ column densities for $\approx 10^{12}M_\odot$ halos \citep{werk14} (blue points and histogram) as well as the COS-inferred density from slab photoionization equilibrium modeling (green points).  {\bf Bottom panels:}  Empirical estimates for the emission-frame surface brightness in the \NII\ 6583\AA.  The solid histograms are calculated using the histogram in the top panel for the \NII\ column and assuming different thermal pressures as \NII\ emission probes pressure for a fixed column.  The green points do the same estimate but use the density measurements of \citet{werk16} from equilibrium photoionization models. The horizontal sensitivity forecasts are calculated in the same manner as in Figure~\ref{fig:OIII}. The grey (black) error bars are the measurements of \citet{2018ApJ...866L...4Z} from stacking SDSS galaxies with $M_* < 10^{10.4} M_\odot$ ($M_* > 10^{10.4} M_\odot$), using an ``average'' virial radius of 250~kpc (300~kpc) to choose the $x$-axis value and the upper limit is shown at $2\sigma$. }
    \label{fig:NII}
\end{figure}

The bottom panel in Figure~\ref{fig:OIII} shows empirical estimates for the emission-frame surface brightness in \OIII\ 5007\,\AA, computed using the histogram/individual points for $\langle \log N_{\rm NIII} \rangle$ shown in the top panel assuming a total oxygen to nitrogen ratio of $7$ ($A_O/A_N \approx 7$; \citealt{asplund09}) and similar fractions of doubly ionized oxygen and nitrogen ($f_{\rm OIII}\approx f_{\rm NIII}$).\footnote{The assumption that two ions trace each other should hold at the factor of two level or better. For $n_e=10^{-3}$\cmminusthree, characteristic of the COS-Halos absorbers, we find $f_{\rm OIII}/f_{\rm NIII} = \{0.9, 0.9, 1.1\}$ at $\log_{10} [T/{\rm K}] = \{4, 4.5, 5\}$, and for $n_e=10^{-2}$\cmminusthree and the same respective temperatures we find $f_{\rm OIII}/f_{\rm NIII} = \{0.9, 0.7, 1.1\}$.  These calculations use the \emph{Q18} background model of \citet{KS19}, but are relatively insensitive to the assumed background.} The solid curves are single pressure models that assume the specified $n_e T$ in units of cm$^{-3}$~K. The $10^6$K virialized gas is required to have $n_e T >10\, $cm$^{-3}$~K to be able to cool in a Hubble time, and cooling is too substantial to be supported by stellar feedback for $n_e T>100\;$cm$^{-3}$~K on the halo scale \citep{mcquinn18}. The inner CGM should harbor even larger pressures. 

The sensitivity of Keck/KCWI is indicated with the magenta horizontal dotted line in Figure~\ref{fig:OIII}.  This horizontal line indicates the $\lambda/\Delta \lambda = 1000$ flux sensitivity of $2 \times 10^{20}~\fergsec$ redshifted to the $z=0.2$ CGM-frame, for which \citet{2010SPIE.7735E..0MM} found a S/N $=10$ detection could be possible when averaging over $600$\,sq.~arcmin in a $15$\,hr observation if the sky could be subtracted to their instrumental specification of $0.01$\%.  This line should be taken as rough guidance but in the ballpark of what has already been achieved.  For example, the VLT/MUSE observations of \citet{2017MNRAS.467.4802F} reach a surface brightness sensitivity of $2000\pm 1000~\fphosr$ to H$\alpha$ with $6$hr observation on target. Predictions for the surface brightness sensitivity of the Dragonfly Telephoto array are somewhat higher with \citet{2019ApJ...877....4L} finding detections of $1000~\fphosr$ are possible. The lower horizontal line is a factor of three times improvement in sensitivity over KCWI, as might be expected for a similar instrument and observing duration on an extremely large telescope (ELT). For example, \citet{2019MNRAS.489.2417A} predict sensitivities of $10^{19-20}~\fergsec$ for the HARMONI spectrograph on the ELT, in accord with this estimate. Such a sensitivity probes $n_e T=1$~K~\cmv.

The green points with error bars in the top panel of Figure~\ref{fig:OIII} are the densities inferred by \citet{werk16} using equilibrium photoionization models as labeled by the right axis. These densities are famously lower than expected from cold gas in kinetic pressure balance with the virialized gas predicted in many models. The intensities inferred from these densities are shown in the bottom panel of Figure~\ref{fig:OIII}. Despite these low densities, the \OIII\ surface brightness we predict at all radii where $N_{\rm NIII}$ is measured is within the reach of being detectable using Keck/KCWI. Absorption systems with the highest predicted signal would be promising targets.   
Additionally, \OIII\ 5007\,\AA\ emission has already been detected by stacking SDSS galaxies with stellar mass below/above $10^{10.4}M_\odot$  \citep{2018ApJ...866L...4Z}, and these measurements are shown by the black/grey markers, respectively.  As the pressure-constraining \OIII\ almost certainly inhabits $10^4$\,K gas, this stack is consistent with these lower densities for both galaxy populations \emph{if} the stacked population shows similar columns to COS-Halos.  It is the case over the decade in stellar mass probed by COS-Halos, there are not strong trends in the columns of ions, and, further, that all ions except for \OVI\ appear to be as abundant in their quiescent galaxies as their star forming ones.  So, our \OIII\ 5007\,\AA\ analysis, using the \citet{2018ApJ...866L...4Z} emission measurements, seems to confirm (with much different assumptions than \citealt{werk16}) that low densities are required.

Figure~\ref{fig:NII} is the same as Figure~\ref{fig:OIII} except for the case of \NII\ 6583\,\AA. Unlike for \OIII\ where we had to infer the column of \OIII\ from \NIII, \NII\ has column density measurements (shown in the top panel). The intensities inferred from the \NII\ column densities at a fixed pressure are an order of magnitude smaller than for \OIII, as are the intensities using the photoionization-inferred densities (green points).  The horizontal sensitivity curves for Keck/KCWI and ELTs are indentical to those for \OIII.  Despite the lower intensities than \OIII, the \NII\ intensity estimates for many of the COS-Halos sightlines are within reach for our ELT sensitivity and perhaps detectable for Keck/KCWI with a deeper stare or adopting less conservative assumptions. Interestingly, the \NII\ 6583\,\AA\ stacked intensity of SDSS galaxies (black/grey points) suggests pressures of $nT >100$\,cm$^{-3}$~K, especially for stellar masses of $>10^{10.4}M_\odot$, in contrast to the $nT \sim 10$\,cm$^{-3}$~K suggested by \OIII\ at  40 kpc. This suggests the stacked \NII\ intensity is weighted towards systems with higher pressures. We caution that this conclusion again comes with the caveat that it requires \NII\ columns similar to those found in COS-Halos, and an order of magnitude larger columns in the stacked systems, while unlikely, would explain the stacked measurements for $nT \sim 10$\,cm$^{-3}$~K.  

Spatially extended \OIII\ has also been detected in a compact Green Pea galaxy with no evidence for AGN activity \citep{2019ApJ...882...17Y}, finding a source frame intensity of $I_{\rm 5007A} \approx 1\times10^5\fphosr$ out to 15 kpc. Using equation~(\ref{eqn:OIII}), this indicates $\left({\langle  n_e T \rangle_{\rm OIII} }/[{10^{-1}\; {\rm cm}^{-3} \times 10^4 {\;\rm K}}] \right)  \left( {N_{\rm OIII}}/{10^{15} \; {\rm cm}^{-2}} \right)\sim 1$.  Measuring an \NIII\ column (to infer the $N_{\rm OIII}$) of these sources in the ultraviolet using a background AGN or down-the-barrel absorption against the host galaxy would constrain the pressure of the \OIII-tracing gas. \OIII's lack of allowed ground state transitions for wavelengths redward of the extreme ultraviolet may also mean our collisional excitation-limit formula is applicable for interpreting extended emission around AGN sources \citep{2017ApJ...835..222S}.

A final interesting optical doublet is \OII\ 3726,~3729\,\AA. Its emission may be detectable near a galaxy as Figure~\ref{fig:f1} shows it only emits significantly at $n_e \gtrsim 10^{-2}$\cmminusthree.  For $N_{\rm OII}=10^{14.5}$\,cm$^{-2}$ and  $n_e= 10^{-2}$\cmminusthree, the 3726\,\AA\ line will have an intensity of $60$ and $350~\fphosr$ at $T=1\times 10^4$ and $2\times 10^4$\,K, respectively.

\subsection{UV Lines from Intermediate Temperature Gas}
\label{ss:BL}

Unlike \OVI, all other UV lines observed with HST/COS are from ions that likely emit in lower temperature gas where the collisional emission depends sensitively on the gas temperature distribution around $hc/[k_B\lambda_X] \sim 10^5$K.  This can be seen in Fig.~\ref{fig:fandsigmav} in the Appendix; the ions responsible for other lines do not exist at $T\gtrsim 10^5$K, whereas collisional excitations are exponentially suppressed at $T\lesssim 10^5$K. This sensitivity to such intermediate temperatures provides the potential to use emission to constrain the amount of gas that is cooling and mixing in halos with virial temperature above $10^5$K.

For a given model that defines the gas mass probability distribution by temperature, $P(T)$, we can calculate the line intensity via
\begin{equation}
    I_{\lambda_{XN}}=  \int \; dT  \; 
    \frac{dN_{XN}}{dT} \; \times \left[{\cal B}_{\lambda X} \frac{\langle \sigma_{\lambda X} v\rangle_T \; n_e }{4\pi} \right];  ~~~~~~~\frac{dN_{XN}}{dT} =  A_{X} f_{XN}(T, n_e) P(T) N_H,
    \label{eqn:IntInt}
\end{equation}
where the leftmost equation is derived by simply breaking up our slab expression for emission (eqn.~\ref{eqn:collem}) into components arising from different temperature gas, $A_{X}$ is the abundance of element $X$ relative to hydrogen, and $f_{XN}$ is the fraction of element $X$ in ionization stage $N$.  We use {\sf Cloudy} \citep{cloudy2017} to calculate $f_{XN}$, assuming gas in collisional ionization equilibrium exposed to an unattenuated \citet[Q18 model]{KS19} background radiation field. The total hydrogen column density, $N_H$, is an overall normalization. We obtain its value by requiring the column in some ion, which we take to be \OVI, to be consistent with observations (see below). Once set, the normalization gives the amount of gas along a line of sight specific to the model, but independent of the ion and transition for which we calculate the emission. In the models we consider, the gas is either isochoric or isobaric, $n_e$ is a function of temperature only, and we do not require a second integral over density to compute the intensity.

We consider two models for the mass-weighted gas temperature distribution $P(T)$:  
 \begin{description}
 \item[Cooling gas model]  This model predicts that the gas mass at each temperature is inversely proportional to the cooling rate, so that $P(T) \propto [\Lambda(T) n_e]^{-1}$.  Almost all isobaric models for intermediate temperature gas follow this distribution at low enough temperatures when the cooling rate is highest.  However, such a relation could hold even up to high temperatures, being condensations from the hot virial gas, potentially explaining the large \OVI\ columns observed by COS-Halos \citep{mcquinn18}. Models that assume the temperature distribution for cooling gas to reproduce the \OVI\ column in COS-Halos do not overpredict the observed columns in other ions but do require a substantial fraction of the halo-associated gas, $\sim$10-100\% by mass, to be participating in cooling \citep{mcquinn18}.\footnote{These models are not cooling flow models where there is extra adiabatic heating from the gas falling deeper into the potential well \citep[e.g.][]{stern19}.}
  
 \item[Turbulent boundary layers and mixing model] Much recent interest in CGM research is in turbulent boundary layers \citep{2019MNRAS.487..737J, 2020ApJ...894L..24F, 2021MNRAS.502.3179T, 2021MNRAS.508L..37T}. While calculations find a single boundary layer is unlikely to contribute much of the \OVI\ column, many boundary layers could add to an appreciable column \citep[e.g.][]{2019MNRAS.487..737J}. Such a situation may occur if the CGM is composed of many small cloudlets \citep{mccourt18}. Several studies have found a flat $T$ distribution by volume\footnote{~Resulting in equal volume per linearly-spaced (rather than logarithmic) temperature intervals.} between the photoionized and virial temperature phases that are mixing such that the mass-weighted gas probability distribution has the scaling $P(T) \propto n_e$ \citep{2019MNRAS.487..737J, 2020ApJ...894L..24F, 2021MNRAS.502.3179T}. A similar flat temperature distribution per unit volume is also found for gas that is turbulently mixing \citep{2022MNRAS.510.3778M}.\footnote{In detail, the gas temperature distribution in boundary layer and other mixing scenarios likely traces cooling at low enough temperatures, with the exact values depending on the strength of turbulent mixing or conduction. At high enough temperatures, when the cooling rate is long compared to the mixing/conduction timescale, there should be an enhancement in the gas reservoir over our flat distribution \citep{2021MNRAS.508L..37T}. The former cooling limit is then our first model, and the latter enhancement would be to modestly increase our \OVI\ intensities relative to the others that owe to lower temperature gas. See \citet{2021MNRAS.508L..37T} for more detailed boundary layer models.} It would be interesting to consider energetics bounds on the amount of mixing gas in such models like have been done for cooling gas models \citep[e.g.][]{mcquinn18}, as bounds will be even more stringent in this scenario.
 \end{description}
 
For each of the cooling and boundary layer/mixing models, we consider two scenarios, where the gas is isobaric -- by which we mean kinetic pressure equilibrium -- or isochoric -- by which we mean constant density.  We use the density parametrization
\begin{equation}
    n_{e} = n_{e,0} \left(\frac{T}{10^{4.3}{\rm ~K}} \right)^{-\gamma},
    \label{eqn:ne0}
\end{equation}
where $\gamma = 1$ for isobaric and $\gamma=0$ for isochoric. Physically, an isobaric distribution seems like the more natural choice, as without invoking some nonthermal bogeyman, the gas tends to cool and condense in this fashion \citep[e.g.][]{mccourt18}. However, an increasingly large contingent of CGM researchers thinks that nonthermal pressure from cosmic rays could be a substantial source of additional pressure and potentially lead to isochoric conditions (or more likely something in between these two limits; \citealt{2018ApJ...868..108B, 2020MNRAS.496.4221J}).  Isochoric conditions could help explain the low densities relative to that expected from isobaric models that are inferred for the cooler CGM clouds \citep{werk14}. Our fiducial density choice, of $n_{e,0} = 10^{-3}$\cmminusthree, is in accord with the median of estimates from equilibrium ionization modeling \citep{prochaska17}, but we also consider $10^{-2}$\cmminusthree that may better reflect the density of the cold clouds in the inner $L_*$-galaxy CGM or in more massive halos.

We can put an upper limit on the amount of emission from intermediate temperature gas by normalizing the emission to the observed \OVI\ column density. The motivation for using the \OVI\ column is that the \OVI\ is most robustly associated with $T\sim 10^{5.5}$~K gas, closest to the intermediate temperatures we are modeling.  Of course, in many models this \OVI\ is thought to be associated with virialized gas, generally requiring a temperature gradient in this phase to produce sufficient $T\sim 10^{5.5}$~K gas \citep{faerman20}. Hence, our predictions will overpredict the emission in models where the bulk of the \OVI\ is virial gas that is not tied to lower temperature gas in the manner we have assumed. In other models, where the \OVI\ owes to cooling, mixing and boundary layers, the emission could be as large as we predict. These models better explain the striking kinematic alignment of cold clouds and the \OVI\ absorbers \citep{werk16}. One might argue our claim that we are putting an upper limit on the emission can be avoided by depressing the amount of $T\sim 10^{5.5}$~K gas relative to the $T\sim 10^{4.5}-10^5$~K emitting gas.  However, we are unaware of models for the gas distribution at intermediate temperatures that have such $P(T)$.

\begin{figure}
\centering
\rotatebox{0}{\includegraphics[width=17cm]{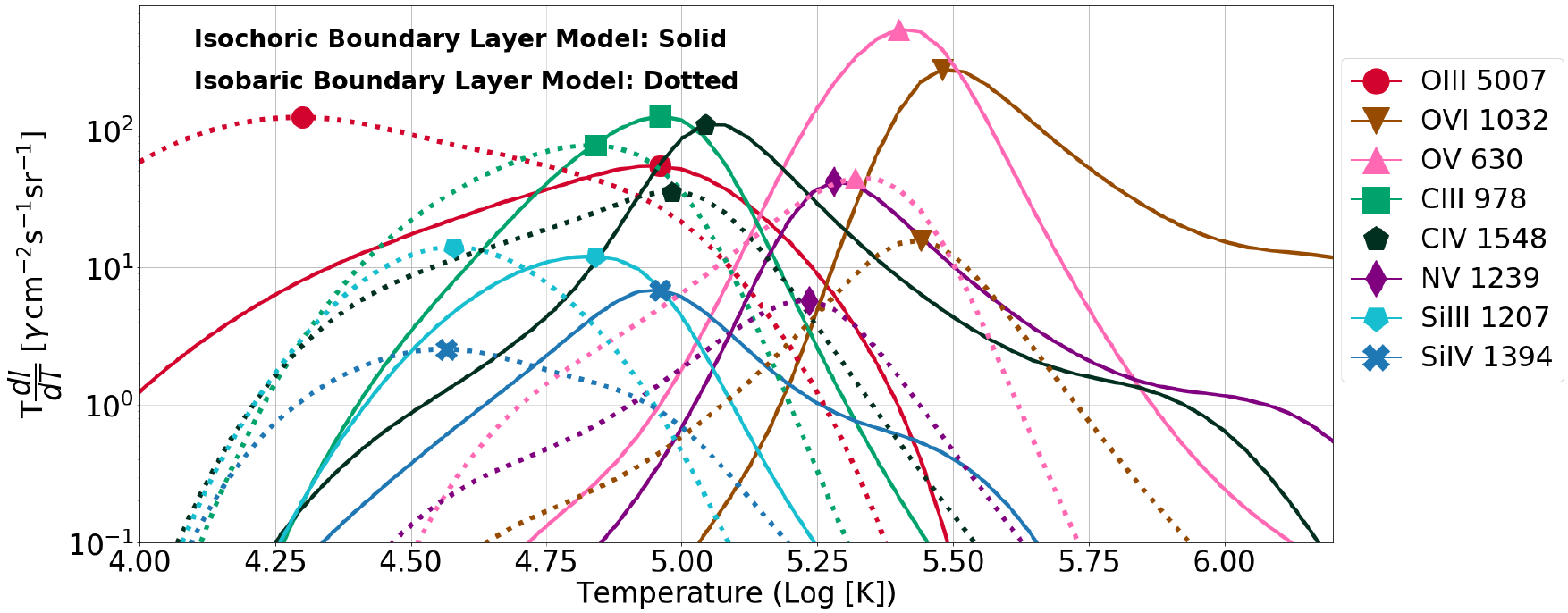}}
\rotatebox{0}{\includegraphics[width=17cm]{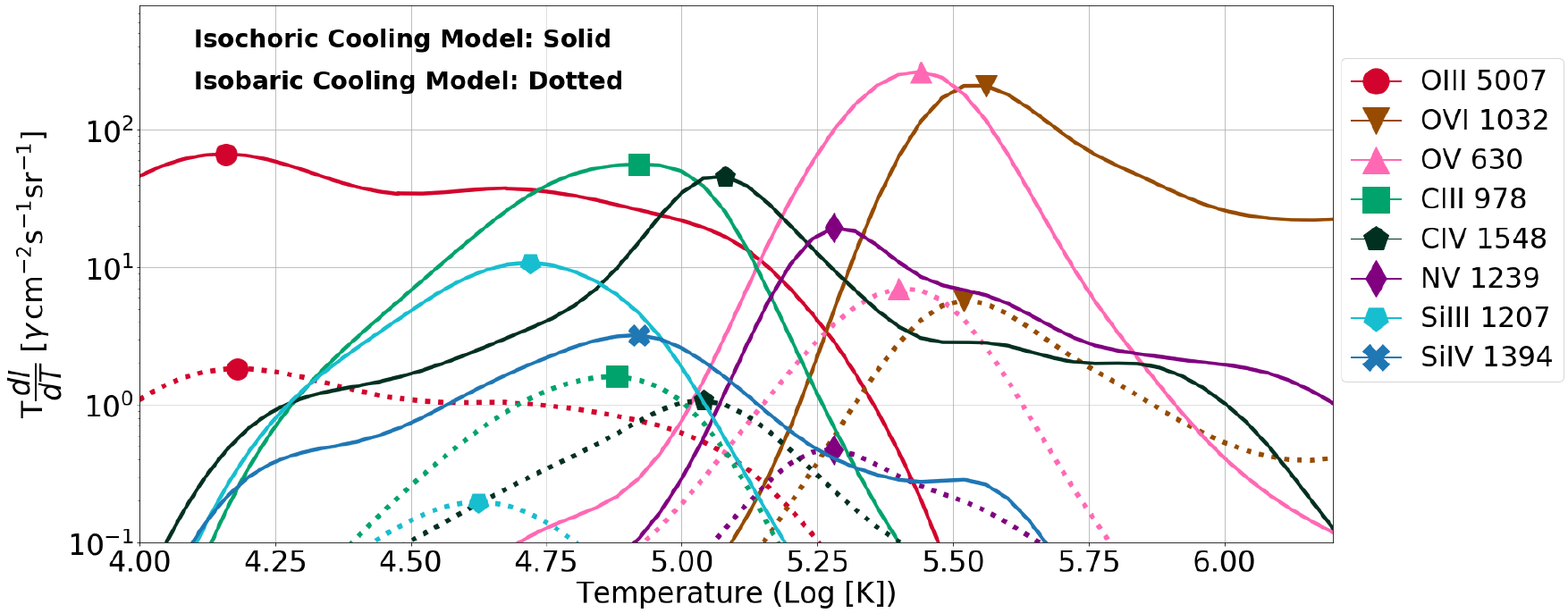}}
\caption{The differential contribution to the line intensity per $\ln(T)$ for $n_{e,0} = 10^{-3}$cm$^{-3}$  (c.f. eqn.~\ref{eqn:ne0}).  Namely, the vertical axis is the product of the integrand in equation~(\ref{eqn:IntInt}) and temperature, such that the linear area under each curve is proportional to the total emission.  All curves are normalized to create an \OVI\ column of $N_{\rm OVI}=10^{14}$cm$^{-3}$, but the results scale linearly with $N_{\rm OVI}$. All line intensity curves are shown with a marker that denotes both the specific spectral line and the intensity peak. {\bf Top panel:}  Models take a flat gas temperature distribution by volume, as motivated by boundary layer and turbulent mixing calculations ($P(T) \propto n_e$).
Essentially the brightest transitions from Table~\ref{table1} have been selected. {\bf Bottom panel:}  Models take the gas temperature distribution to be that of gas cooling freely  ($P(T) \propto [\Lambda(T) n_e]^{-1} $). The same bright transitions as the top panel are shown, although \SiIV\ 1394\,\AA\ falls below the minimum shown.}
\label{fig:mixing&cooling}
\end{figure}

Figure~\ref{fig:mixing&cooling} shows the differential contribution to the intensity per $\ln T$, $dI_{\lambda_X}/d\ln T$, for the most emissive transitions from Table~\ref{table1}, for the case where $n_{e,0} =10^{-3}$\cmminusthree and normalized such that $N_{\rm OVI} = \int dT \, dN_{\rm OVI}/dT = 10^{14}$\,cm$^{-2}$.\footnote{This results in a total hydrogen column density of $N_H \approx 10^{19.1}$\cmc~for the isochoric boundary layer and isobaric cooling models, and $N_H \approx 10^{20.4}$\cmc~for the isobaric boundary layer and isochoric cooling assuming a maximum temperature of $10^6$K. These values can change by a factor of up to about two when the upper limit of the temperature range is varied between $10^{5.5}$ and $10^{6.5}$\,K.} This $N_{\rm OVI}$ is a factor of a few lower than the median column for $L_*$ galaxies \citep{tumlinson11, werk14}, and as shown in equation~\eqref{eqn:IntInt}, the emission scales linearly with the \OVI\ and total gas density column. In fact our estimates may have wider applicability as columns of $N_{\rm OVI} = 10^{14-15}$\,cm$^{-2}$ are observed for star forming galaxies over a broad range in stellar mass \citep{kirill}. With this normalization, we have checked that the columns associated with other ions in all of our $P(T)$ models are below measured values; the columns that are observed from absorption studies in other ions would be associated with a pile-up of gas at the equilibrium temperature of $\sim 10^4$~K, gas which does not emit substantially in the UV.

The top and bottom panels show our `mixing' models with $P(T) \propto n_e$ and cooling models with $P(T) \propto [\Lambda n_e]^{-1}$, respectively. The linear areas under the curves per logarithmic temperature interval are proportional to the total emission in each species, and each curve's span in temperature illustrates the range of temperatures where each ion contributes to collisional emission. Most UV transitions are most emissive at $10^{4.5}-10^5$\,K.  Exceptions are \OV\ 630\,\AA\ and \OVI\ 1032\,\AA, which emit at $10^{5}-10^{5.75}$\,K.\footnote{ That \OVI\ 630\AA\ emits at just slightly lower temperatures \OVI\ 1032\,\AA\ suggests that it is a good test of any model for the origin of the large \OVI\ columns and not just the models considered here.} The isochoric scenario (solid curves) produces more emission at lower densities than the isobaric (dotted) since the gas is denser at the temperatures where emission peaks. The amount of emission is higher for most transitions in the mixing model as the cooling model results in less gas at the intermediate temperatures where the cooling rate peaks. 

\begin{figure}
\centering
\rotatebox{0}{\includegraphics[width=17cm]{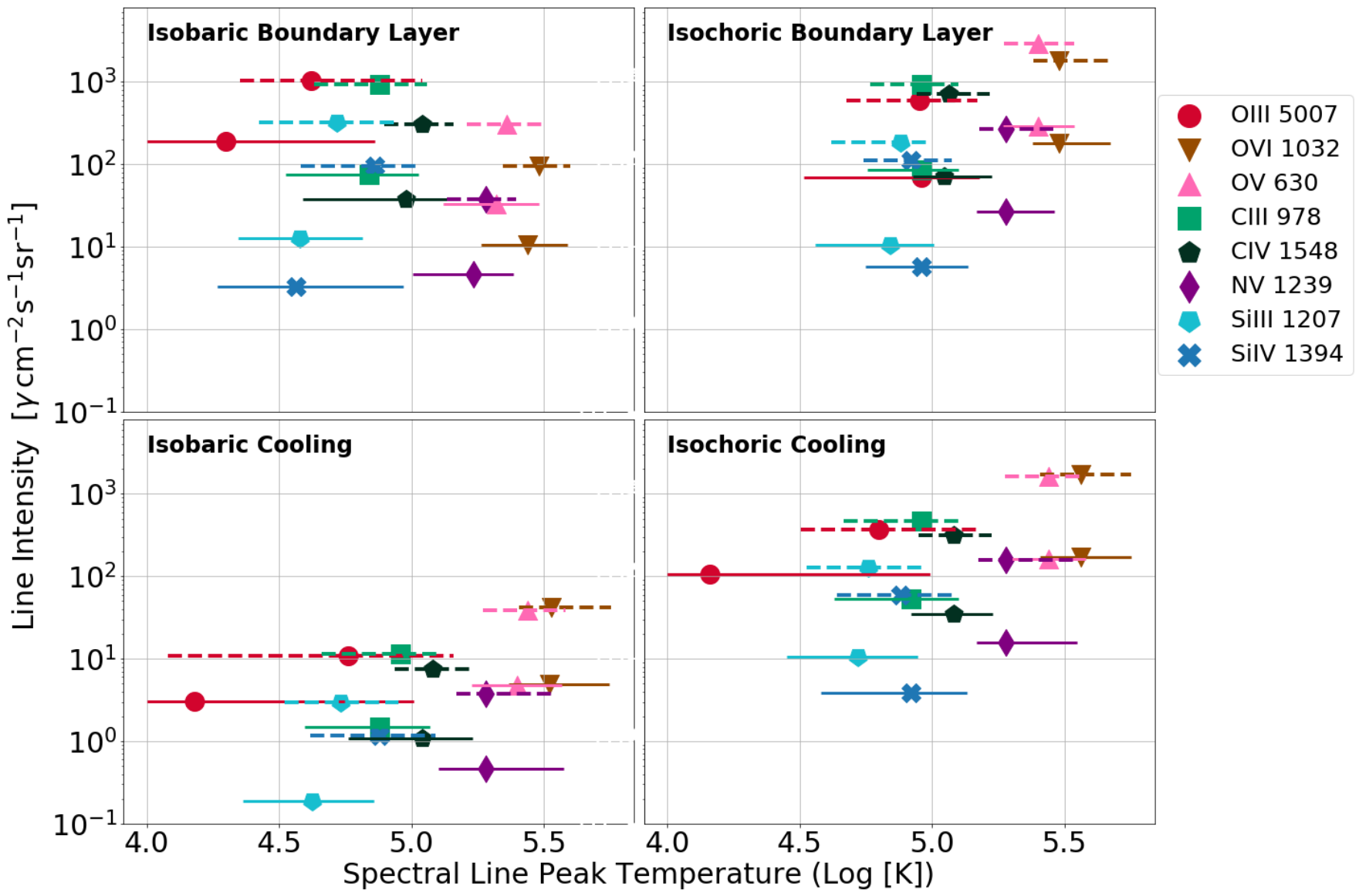}}
\caption{Total line intensity in the boundary layer and cooling models, as calculated by equation~(\ref{eqn:IntInt}) and normalized to create an \OVI\ column of $N_{\rm OVI} = 10^{14}$cm$^{-2}$, but the results scale linearly with $N_{\rm OVI}$.  The markers denote the temperature at which $dI_{\lambda_X}/d\ln T$ peaks and the bars indicate the temperature range where the intensity is greater than one third of the line's peak intensity, $dI_{\lambda_X}/d\ln T > \max[dI_{\lambda_X}/d\ln T]/3$. For each model, two pivot densities are shown as defined by equation~(\ref{eqn:ne0}), $n_{e,0}=10^{-3.0}$\cmminusthree (solid bars) and $n_{e,0}=10^{-2.0}$\cmminusthree (dashed bars). 
\label{fig:comparisons}}
\end{figure}

Figure~\ref{fig:comparisons} shows the total intensities of the most emissive lines (the integrals over the curves in Fig.~\ref{fig:mixing&cooling}), with the horizontal error bars indicating the ranges in gas temperature that contributes most to the emission, with the range given by $dI_{\lambda_X}/d\ln T > \max[dI_{\lambda_X}/d\ln T]/3$. We plot the results for the boundary layer model (top panels) and the cooling model (bottom panels), with the isobaric and isochoric scenarios (left and right panels, respectively). The dashed lines show the case of denser gas with $n_{e,0} = 10^{-2}$\cmminusthree, in addition to our fiducial value of $n_{e,0} = 10^{-3}$\cmminusthree shown with the solid. We find that \OVI\ 1032\,\AA\ and \OV\ 630\,\AA\ are the strongest lines for the isochoric models.  Additionally, \CIII\ 978\,\AA\ and \CIV\ 1548\,\AA\ are the strongest lines from gas at temperatures of $10^{4.5} -10^{5}\,$K.  For all but the isobaric cooling model (bottom left), the predicted intensities in these two lines are above the threshold of what is observable with a $1\;$m UV space telescope ($\sim 100\;\fphosr$) in the inner $\sim 0.2\,r_{\rm vir}$ CGM (eqn.~\ref{eqn:sens}). This indicates that such a sensitivity would test all models except the isobaric cooling one. For $T<10^{4.5}$\,K, the optical \OIII\ 5007\,\AA\ transition is the most promising line. However, \OIII\ 5007\,\AA\ and other optical transitions are likely not useful for testing models for intermediate temperature gas since they emit substantially in the $10^4$\,K equilibrium phase (\S~\ref{ss:optical}).

The intensity estimates in this section do not depend on the gas metallicity for the boundary layer/turbulent mixing model. For the cooling gas model, these estimates are approximately independent of the metallicity, as long as the gas radiative cooling is dominated by metal ions.  We have used the $\Lambda(T)$ for $Z = 0.3~\Zsun$, but the independence of our results on metallicity should hold to $Z \approx 0.1~\Zsun$, below which \HeII~cooling starts to dominate and the metal emission would be reduced in proportion to the fractional contribution to cooling of helium.

\subsection{Hydrogen and Other Recombination Emission}
\label{sec:resultsHI}
\label{ss:HI}

As highlighted previously, \HI\ is likely to be the only ion that is detectable in recombination emission off the ionizing background, particularly given our restriction to ions with ionization potentials to get to their stage of $>1$\,Ry. \HI\ recombination emission could potentially be observed either via its EUV Lyman transitions or optical Balmer transitions.

The metagalactic \HI\ ionizing background is well constrained by the Lyman-$\alpha$ forest to have a value of $\Gamma_{\rm HI, bkgd} = 4\times10^{-14} (1+z)^5$\,s$^{-1}$ at $z\lesssim 2$ \citep{2015ApJ...811....3S}.  Using this rate, we can rewrite equation~(\ref{eqn:photoX}) for the CGM-frame intensity in recombinations from photoionized \HI\ as 
\begin{eqnarray}
I_{\rm REC, HI}(n'\rightarrow n)
&\approx& 1.0\times10^3  (1+{\cal E})  \left( \frac{f_{n'n}}{0.2} \right) \; \left(1- \exp\left[-\frac{N_{\rm HI}}{10^{17.8} {\rm cm}^{-2}}\right] \right) \times \left(\frac{1+z}{1.3} \right)^5~\fphosr , 
\label{eqn:photoHI} 
\end{eqnarray}
where $f_{n'n}$ is the number of recombinations that transition through the $n'\rightarrow n$ radial quantum numbers (e.g. $f_{32}\approx 0.15$ for H$\alpha$), and ${\cal E}$ is the enhancement factor from host galaxy emission over the background-sourced emission that is estimated in the following footnote.\footnote{In the optically thin limit where $N_{\rm HI} \ll 10^{17.8}$cm$^{-2}$, this enhancement is given by
\begin{equation}
    {\cal E}\big |_{\rm op.thin} = \frac{\sigma_{\rm eff, *} f_{\rm esc} d\dot N_\gamma/d M_\star ~ {\rm SFR}}{\Gamma_{\rm HI, bkgd} \times 4\pi r^2} = 0.4 \left( \frac{f_{\rm esc}\, {\rm SFR}}{0.1\; M_\odot\;  {\rm yr}^{-1}} \right) \left( \frac{r}{100 {\rm \, kpc}} \right)^{-2} \left(\frac{1+z}{1.2}\right)^{-5},
    \label{eqn:prox}
\end{equation}
where $d\dot N_\gamma /d M_\star = 2\times10^{53} {\rm ~\gamma ~s^{-1}} (M_\odot {\rm yr}^{-1})^{-1}$ is the number of ionizing photons produced per stellar mass formed \citep[e.g.]{Topping_2015}, and where a stellar-like spectral index of $-2$ to compute the spectrum-weighted photoionization cross section $\sigma_{\rm eff, *}$. For typical star formation rates of Milky Way-like galaxies and $\gtrsim 100\, $kpc impact parameters, ${\cal E} \gg 1$ only if the escape fraction of ionizing photons, $f_{\rm esc}$, is appreciable for a particular galaxy.\label{foot:prox}} To evaluate $\sigma_{\rm eff}$, we have assumed an ionizing background spectrum with spectral index in specific intensity of $-1$ as motivated by ionizing background models (see \S~\ref{foot:sigeff}).

\begin{figure}
    \centering
    \includegraphics[width=15cm]{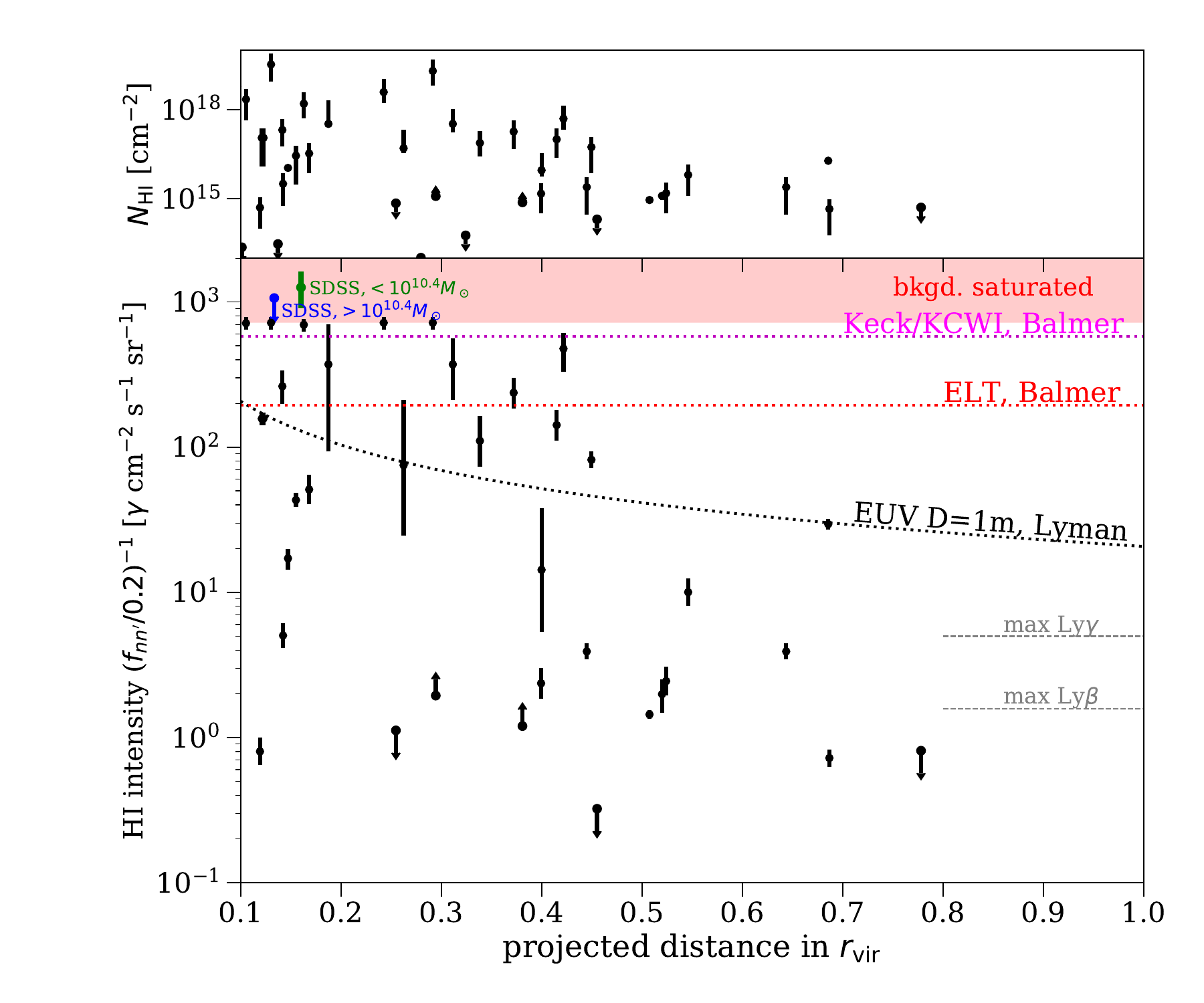}
    \vspace{-.3cm}
    \caption{ {\bf Top panel:} COS-Halos constraints on the \HI\ column density around $\approx 10^{12}M_\odot$ halos \citep{prochaska17}.  {\bf Bottom panel:}  Estimated \emph{CGM-frame} intensity from fluorescence from the extragalactic ionizing background using equation~(\ref{eqn:photoHI}). Estimates ignore the host galaxy's proximate contribution, which may significantly enhance the emission, particularly in the inner CGM (c.f. eqn.~\ref{eqn:prox}). In both panels, the different colors differentiate limits and detections.  The pink region denotes the saturating intensity from self-shielding, and the dashed horizontal line segments on the right-hand side are the maximum emission anticipated in Lyman lines owing to absorptions. The dotted horizontal forecasts for the sensitivity of optical instruments are calculated in the same manner as in Figure~\ref{fig:OIII}, and the black dotted curve shows the sensitivity of the $D=1\,$m EUV satellite for the same specification as in Fig.~\ref{fig:OVI}.  The labeled error bars are the measurements of \citet{2018ApJ...866L...4Z} from stacking SDSS galaxies with stellar masses above and below $10^{10.4} M_\odot$ at $40\,$kpc and scaled to a $z=0.2$ emission-frame, with the upper limit shown at $2\sigma$.}
    \label{fig:HI}
\end{figure}

The \HI\ column density, $N_{\rm HI}$, is typically in the range $10^{15}-10^{18}$\,cm$^{-2}$ for $z\approx 0.2$ Milky Way CGMs on $100~$kpc scales \citep{tumlinson11, prochaska17}, although CGM \HI\ columns can become considerably larger with increasing redshift. Figure~\ref{fig:HI} shows emission-frame intensity estimates for $z\approx 0.2$ Milky Way-like galaxies, using equation~(\ref{eqn:photoHI}) with no proximate enhancement (${\cal E}=0$), and using the $N_{\rm HI}$ measurements of the COS-Halos sample reported in \citet{prochaska17}. The black points with errorbars in the top panel show the measured columns and the bottom the background-sourced source-frame intensity calculated using equation~(\ref{eqn:photoHI}).  As a significant fraction of these points are above the detection thresholds of the KCWI and an IFU on an ELT -- thresholds that should be taken as rough guideposts --, this suggest that some \HI\ systems have detectable H$\alpha$ surface brightnesses.  In support of these guideposts, the VLT/MUSE analysis of \citet{2017MNRAS.467.4802F} achieved sensitivity of $2000\pm 1000~\fphosr$~to H$\alpha$ in a $6\,$hr observation.  This background-sourced  signal will increase for systems at higher redshifts, with the observed photon number intensity $I_\nu$ scaling as $\approx (1+z)^2$.  

Let us focus first on Hydrogen Balmer emission  ($n =2$), where the brightest line is H$\alpha$.  Collisional emission is subdominant, as detailed in the ensuing footnote.\footnote{Collisional excitation emission of H$\alpha$ is smaller than recombination emission for \HI\ fractions less than $10^{-2}$ at $T>20,000$K \citep[see their table 11.5]{2006agna.book.....O}, and the  \HI\ fractions have to be even larger at lower temperatures.  We can calculate the \HI\ fraction from the photoionization rate $x_{\rm HI} = \alpha_A n_e/\Gamma_{\rm HI} =2\times10^{-4} (T/20000{\rm \, K})^{-0.7}\times  (n_e/10^{-3} {\rm cm}^{-3})  \times (10^{-13}~{\rm s}^{-1}/[(1+{\cal E})\Gamma_{\rm HI}])$. Since the densities of the low redshift Milky Way CGM are likely well below the  $n_e \approx 10^{-1} {\rm cm}^{-3}$ required to reach $x_{\rm HI} \approx 10^{-2}$, we can safely ignore collisional excitations for gas that is not substantially self-shielding.    The same conclusion is even stronger for other Balmer lines and also holds for the $n'>2$ Lyman-lines since the excitation process is the same.}  We can compare our predicted intensities to the SDSS stacked measurements centered at $40\,$kpc of \citet{2018ApJ...866L...4Z}, which are shown with the blue upper limit and green error bars for galaxies above and below stellar masses of $10^{10.4}M_\odot$, respectively.  These observer-frame intensities have been scaled to $z=0.2$ emission-frame intensity. The green is near the prediction for the background-saturated signal. \citet{2018ApJ...866L...4Z} also reported a measurement centered at 17~kpc, of $\approx 5000~\fphosr$ for each stellar masses bin. Assuming the \HI\ column is near that needed for the signal to saturate, their measurement at 17~kpc requires a modest proximate enhancement; if 1\% of the ionizing flux escapes for  ${\rm SFR} = 1 M_\odot$\;yr$^{-1}$ this would result in a factor of three enhancement at $17\;$kpc (c.f. footnote~\ref{foot:prox}).

Next consider the Lyman series ($n =1$). While Ly$\alpha$ ($n'=2$) is not suppressed by radiative transfer, at least if one ignores dust absorption, emission from the host galaxy that scatters out into the CGM likely far exceeds the Ly$\alpha$ luminosity of the CGM. Higher $n'$ Lyman lines scatter less and so their CGM emissions should not be as contaminated by emission from the host galaxy. However, the scattering that they experience destroys these photons, diminishing the potential CGM intensity. For Ly$\beta$ ($n'=3$, 1026\,\AA), 12\% of the time scattering leads to the Ly$\beta$ photon being destroyed rather than re-emitted (e.g.~\citealt{pritchard}), suggesting that for systems with $\tau_{{\rm Ly}\beta} \gg 10$, the emission is from the surface of the CGM, rather than its volume as for the other lines we consider. For $n'>3$, destruction occurs with $\approx 20\%$ probability at each scattering \citep{pritchard}, but the cross section for absorption is also smaller. Ly$\beta$ and Ly$\gamma$ emission becomes surface for column density thresholds of $N_{\rm Ly\beta, crit} = 1\times10^{15} (b/20\; {\rm km \; s^{-1}})$\;cm$^{-2}$ and $N_{\rm Ly\gamma, crit} = 3\times10^{15} (b/20\; {\rm km \; s}^{-1})$\;cm$^{-2}$, where $b$ is the linewidth and both columns are defined by requiring they yield a line-center optical depth of ten. As these critical columns are almost a thousand times below the saturation column of $N_{\rm HI} \approx 10^{17.8}$\;cm$^{-2}$ in equation~(\ref{eqn:photoHI}), the photoionization driven surface brightness for Ly$n'$ radiation $ n' >2$ saturates at a CGM-frame intensity of $I_{\rm REC, HI} \sim 1  (1+{\cal E}) f_{n'n} (n'/3)^3 ~\fphosr$. The small $I_{{\rm Lyn}', {\rm crit}}(3\rightarrow 1)$ suggests that Ly$\beta$ 1026\,\AA\ should not be a worry to contaminate one of the most promising lines \OVI\ 1032\,\AA, at least for foreseeable sensitivities and if the proximate enhancement is not large.

We can perform similar estimates for the fluorescent intensity for helium and metal lines, but unless near a bright source their intensities are likely too small to be observable. Let us first consider helium. As shown in \S~\ref{sec:energetics}, \HeII\ will be optically thick for systems with $N_{\rm HI}\gtrsim10^{16}$\,cm$^{-2}$ and hence largely in \HeII, at least up to damped Lyman-$\alpha$ systems at which point the helium becomes \HeI.  We find that even without self shielding the weak $z\approx 0.2$ \HeII\ ionizing background in ionizing background models is only able to keep the \HeII\ half ionized for our COS-Halos like reference density of $n_e=10^{-3}$\cmminusthree, so the helium will quickly become \HeII\ if it is not already once self-shielding occurs.  Self-shielding means that the column of \HeII\ will be above the critical column for which the background-fluorescent emission is saturated.  Ionizing background models predict \HeII\ photoionization rates of $\Gamma_{\rm HI}/ \Gamma_{\rm HeII}\approx 100$ at $z\lesssim 3$ \citep{faucher}, resulting in factor of a hundred lower saturated intensities in fluorescence compared to hydrogen, as given in equation~(\ref{eqn:photoX}). Since self shielding results in the bulk of the helium being in \HeII, some of the \HeII\ will recombine into \HeI; recombination lines of \HeI\ are more promising than of \HeII.  Since \HeI\ has a similar recombination rates to that of hydrogen and since ionizing background models predict a photoionization rate that is only a factor of two smaller \citep[e.g.]{KS19}, the 7\% helium density by number means that \HeI\ lines will saturate at a similar intensity to hydrogen but for $N_{\rm HI}$ that are about a factor of ten higher.  Only a small fraction of COS-Halos sightlines show high enough $N_{\rm HI}$ for \HeI\ emission to saturate, and the \HI\ columns required for saturation will also result in the \HeI\ EUV recombination radiation being absorbed by the \HI. Absorption aside, once $N_{\rm HI}>10^{16}$\,cm$^{-2}$ such that the \HeII\ self shields, the intensities in \HeI\ recombination lines are similar to those found in equation~(\ref{eqn:photoHI}) for hydrogen, but with the characteristic hydrogen column $10^{17.8}$\,cm$^{-2}$ increased by an order of magnitude.  This means that the predicted intensity is a factor of ten or so smaller than hydrogen H$\alpha$ for most columns. Thus, while difficult, \HeI\ recombination emission may not be impossible to detect.  The prospects for metal line recombination emission are more dismal. Metal lines from species that exist in abundance when the hydrogen is ionized are not only optically thin to continuum photons, with the largest columns likely reaching $N_X\sim 10^{15}$\,cm$^{-2}$ as suggested by the COS-Halos observations, but also their photoionization rates are generally much smaller than that of hydrogen as their lowest bound-free transition is more energetic. Roughly, the maximum recombination intensity in the line $\lambda_X$ should be smaller compared to the saturated hydrogen intensity by the factor $N_X/( 10^{17.8}\,{\rm cm}^{-2})\times \Gamma_{\rm X}/\Gamma_{\rm HI}$.  Thus, in the absence of a large proximate enhancement, an observable line intensity from ions with $E_{{\rm ion}, X} >1$~Ry must be driven by collisions.

\section{conclusions}

This work investigated the feasibility of detecting various UV and optical emission lines, tying our estimates to the observed ionic column densities from HST/COS absorption studies.  Although CGM line emission is dim, our work identified several emission lines that may be observable with existing or future instruments. Our results include:
\begin{itemize}
\item We identified the most emissive lines at temperatures within the range of $10^4 - 10^6$\,K and at densities relevant for the CGM.  Several ions can emit ten percent or more of the total luminosity at a given temperature, including \OIII\ at $10^{4-4.5}$K; \CIII\ and \CIV\ at $10^{4.5-5}$\,K; \OV\ and \OVI\ at $10^{5-5.5}$\,K.

\item The emission in \OVI\ $1032, 1038$\,\AA\ is insensitive to temperature over the range of temperatures where \OVI\ is present in collisional ionization equilibrium. This means that, if the typical \OVI\ column is known from absorption measurements, a measurement of \OVI\ in emission constrains the gas density. We used models for the density of virialized gas to predict the intensity, showing that \OVI\ emission in the inner 100 kpc of Milky Way CGMs is within reach of a $D=1\;$m space telescope.  We also commented on the implications of existing \OVI\ CGM emission measurements at $\sim 10\, $kpc impact parameters from the host galaxy.

\item The emission in optical lines of \OII, \OIII, and \NII\ is sensitive to the $\sim 10^4$\,K phase probed by most UV absorption lines. We found that, owing to the linear scaling with temperature of their collisioinal excitation rates over the temperature range these ions can inhabit, \OIII\ 5007\,\AA\ and \NII\ 6583\,\AA\ constrain the pressure of the CGM. We used existing stacked measurements of SDSS galaxies \citep{2018ApJ...866L...4Z} to show that, if these galaxies have similar columns as in the COS-Halos sample, the \OIII-tracing gas pressures seemed consistent with the low inferred pressures of \citet{werk16} from photoionization modeling.  However, for instead their $>L_*$ galaxy stack of \NII\ 6583\,\AA, our inferred pressure would be higher than \citet{werk16} under the same assumption.  We further found that \OIII\ 5007\,\AA\ and \NII\ 6583\,\AA\ emission out to $\sim 100$\,kpc should be within the reach of existing IFUs on $\approx 10$\,m-class telescopes.

\item Most UV emission lines are sensitive to gas at temperatures $10^{4.5}-10^{5.5}$\,K, as at higher temperatures the ions do not exist, whereas at lower temperatures they cannot be excited collisionally. The amount of such intermediate temperature gas is not well constrained, with some simulations suggesting that it could be substantial. We developed models for the amount of $10^{4.5}-10^{5}$\,K gas that were motivated by simulations of turbulent boundary layers/mixing gas and cooling.  These models were normalized to produce a given \OVI\ column. We found that in many scenarios where an appreciable fraction of the observed \OVI\ column in COS-Halos owes to such processes, some ions emit sufficiently to be detectable with a $D=1$m UV space telescope. Some of the most promising lines are \OV\ 630\,\AA, \CIII\ 978\,\AA, and \CIV\ 1548\,\AA. Indeed, \OV\ emits substantially at $10^{5}-10^{5.5}$\,K and will be the brightest CGM emission line if much of the \OVI-bearing gas resides at $10^{5.5}$\,K.

\item \HI\ Balmer emission is the most detectable line in recombination.  We showed that the ionizing background can drive an H$\alpha$ intensity as large as $\approx 10^3 [(1+z)/1.5]^5\fphosr$, where reaching this maximum requires $N_{\rm HI} \gtrsim 10^{17.8}$\,cm$^{-2}$, which is satisfied for less than a third of COS-Halos sightlines. Higher Lyman-series recombination radiation is unobservably small owing to radiative transfer effects, as are recombination photons from metal ions or \HeII\ that are driven by fluorescence off the UV background. Existing stacked H$\alpha$ measurements at $40\;$kpc are consistent with the maximum intensity from background fluorescence, and smaller radii show evidence for a proximity effect. \HeI\ recombination lines should be an order of magnitude smaller than \HI\ Balmer for most COS-Halos sightlines, modulo transition probabilities.

\item A glaring omission from our investigation is the famous \HeII\ 1640\,\AA\ Balmer line. However, we can apply some of our results to understand its emission. First, it will be undetectable in background-driven recombination radiation because the \HeII\ photoionization rate is so low, and so like most ions we considered is only detectable in collisional emission. Over the temperature range where \HeII\ drives the collisional cooling for primordial gas, $2-4\%$ of its emission comes out in this line \citep{2006ApJ...640..539Y}. Equation~(\ref{eqn:Ienergy}) suggests that $2-4$\% is sufficient to be detectable if a significant fraction of feedback energy is radiated in the CGM, with this percentile just somewhat smaller than the most promising metal lines. However, at our fiducial metallicity of $0.3 Z_\odot$, the fraction of cooling by primordial gas is down by 1-2 orders of magnitude at $10^5-10^6$\,K and so we anticipate only a fraction of a percent of the energy comes out in the 1640\,\AA\ line at this metallicity. Consequently, \HeII\ 1640\,\AA\ emission is a more promising target for low metallicity CGM gas.

\end{itemize}

Our work follows in the footsteps of \citet{2013MNRAS.430.2688V} and \citet{2016ApJ...827..148C}.  Both used numerical simulations and, like us, \citet{2016ApJ...827..148C} used observed ionic column densities. These studies identified similar lines as being the most promising metal lines, including \CIII\ 978\,\AA\, \CIV\ 1548\,\AA\, and \OVI\ 1032\,\AA.  Indeed, \citet{2013MNRAS.430.2688V} found \CIII\ 978\,\AA\ to be their most emissive metal line, which interestingly indicates a substantial reservoir of intermediate temperature gas in their simulation. The predictions of both studies for \OVI\ intensities are below ours for the likely range of densities, which is likely because these simulations underpredict the \OVI\ columns relative to observations. The intensity estimates tied to empirical columns in \citet{2016ApJ...827..148C}, which considered only UV lines, will be low because the $\sim 10^4$\,K equilibrium phase they assumed is too cold to emit substantially. Consistent with our results, \citet{2013MNRAS.430.2688V} and \citet{2019ApJ...877....4L} found H$\alpha$ to be promisingly bright with detectable intensity of $\sim 1000~\fphosr$ out to $\sim 40$~kpc for simulated Milky Way mass galaxies. Their predicted background-fluorescent intensity does sometimes significantly exceed the maximum possible because their simulations do not include self-shielding properly. Our formalism naturally accounts for self shielding.

Our absorption-tied emission estimates can guide integration times on existing instruments and the design of future ones. The tabulation of important cross sections in Table~\ref{table1} may facilitate emission estimates from simulations, without the need for tabulating a grid of photoionization models. \\

DP and ES are co-lead authors of this work. We thank Sarah Tuttle, Tom Quinn, and Jess Werk for helpful conversations. We thank G. Mark Voit for his precipitation models, and Jess Werk, Sarah Tuttle, and Brent Tan for comments on the manuscript. We thank Michele Fumagalli, Dylan Nelson, Huanian Zhang, and the anonymous referee for their suggestions and comments on the manuscript. We acknowledge support from NSF award AST-2007012 and NASA award 19-ATP19-0023. We also wish to acknowledge the scholarship support provided by the Mary Gates Endowment for this project. The collision strengths presented in Table 1 were calculated using CHIANTI, which is a collaborative project involving George Mason University, the University of Michigan (USA), University of Cambridge (UK) and NASA Goddard Space Flight Center (USA).

\section*{Data Availability}
The fractional line intensity data presented in Figure 1, calculated using {\sf Cloudy}, are available in the online supplementary material. In addition to the data we present, we provide the specific fractional emissions as functions of temperature and density for different photoionizing backgrounds.

\bibliography{cgm}{}
\bibliographystyle{aasjournal}

\appendix

Table \ref{table1} provides information on the likely significant metal emission lines for the circumgalactic medium. This appendix will cover the steps to go from this table to CGM emission estimates. The photon number intensity and emissivity are given by
\begin{equation}
    I_{\rm CE} =   \frac{1}{4\pi} {\cal B}_{\lambda_X} \langle \sigma_{\lambda_X} v \rangle_T n_e N_{X}; ~~~~
    \epsilon_{\rm CE} = {\cal B}_{\lambda_X} \langle \sigma_{\lambda_X} v \rangle_T   n_e n_{X},
\label{eqn:inten&emis}
\end{equation}
where the intensity equation repeats equation~(\ref{eqn:collem}) in the main text, after which also follows most of the relevant definitions for understanding equation~(\ref{eqn:inten&emis}).

Table \ref{table1} provides both the ${\cal B}_{\lambda_X}$ and the $\Upsilon/g_{\rm grd}$ terms for the metal emission lines we consider. Here, $\Upsilon$ is the collision strength and $g_{\rm grd} = 2J+1$ is the ground state degeneracy of the ion -- not always the lower energy of the transition -- where $J$ is the total angular momentum quantum number. For a specific spectral line in a gas at temperature $T$, $\langle \sigma_{\lambda_X} v \rangle_T$ can be determined from $\Upsilon/g_{\rm grd}$ \citep{2006agna.book.....O}:
\begin{equation}
    \langle \sigma_{\lambda_X} v \rangle_T = \frac{\beta}{\sqrt{T}} \frac{\Upsilon}{g_{\rm grd}} \exp{\left[\frac{-E_{X}}{k T}\right]};
\end{equation}
\begin{equation}
    \beta = \left( \frac{2 \pi \hbar^{4}}{k  m_e^{3}}\right)^{1/2}  = 8.629 \times 10^{-6} ~~ {\rm cm}^{3}~{\rm s}^{-1}~ {\rm K}^{-1/2}.
\end{equation}

\begin{figure}
\centering
\rotatebox{0}{\includegraphics[width=17cm]{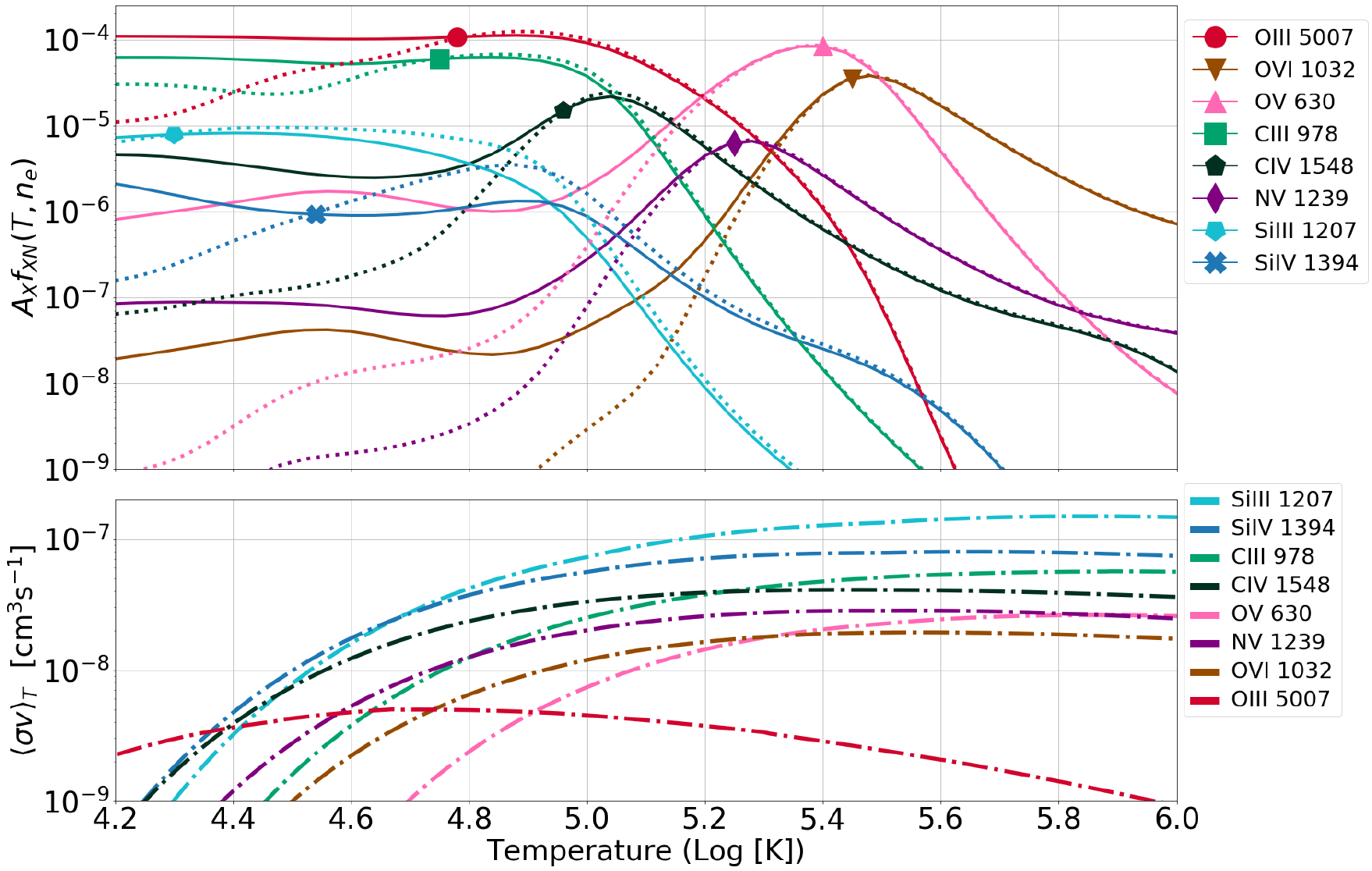}}
\caption{ {\bf Top panel:} Product of an element $X$'s Solar abundance and the fractional abundance in the $N^{\rm th}$ ionization stage ($A_X f_{XN}(T, n_e)$) versus temperature, featuring the most promising transitions from this study. Two densities are shown, $n_e=10^{-3.0}$\cmminusthree, which is represented by solid lines, and $n_e=10^{-2.0}$\cmminusthree, which is represented by the dashed lines. 
{\bf Bottom panel:} The temperature dependence of each transition's $\langle \sigma_{\lambda_X} v\rangle_T$.  That $\langle \sigma_{\lambda_X}  v\rangle_T$ is flat at $T\gtrsim 10^5$K, but strongly declining at lower temperatures, combined with these ions only existing at $T\lesssim 10^5$K as shown in the top panel, means that UV emission is sensitive to the amount of $\sim10^5$K gas.  Optical transitions like \OIII\ probe lower temperatures, and \OV\ and \OVI\ are the ions sensitive to the highest temperatures of $T\sim 10^{5.5}$\,K.
\label{fig:fandsigmav}}
\end{figure}

\section{Calculating emission from collision strengths} 
\label{ap:linestrengths}

The bottom panel in Figure~\ref{fig:fandsigmav} shows $\langle \sigma_{\lambda_X} v \rangle_T$ as a function of temperature for the most intense transitions studied in this paper.  Note that  $\langle \sigma_{\lambda_X} v \rangle_T$ flattens above $\sim 10^5$\,K for the UV transitions, as many electrons in the thermal bath are able to collisionally excite the transition. However, owing to the lower energy of optical transitions (such as \OIII\ 5007\,\AA), their $\langle \sigma_{\lambda_X} v \rangle_T$ peaks at lower temperatures. Indeed, the quasi-linear scaling seen in this plot at $T\lesssim 10^{4.7}$\,K for the \OIII\ 5007\,\AA\ collision rate directly leads to our conclusion that \OIII\ is sensitive to pressure.

Equation~(\ref{eqn:inten&emis}) shows that emission is not only set by the collision rate, but also by the density/column density in an ion. The top panel in Figure~\ref{fig:fandsigmav} shows the fractional abundance of the most abundant elements in ionization states that emit substantially in the optical or UV, $A_X f_{XN}$. This abundance rapidly decreases above $\sim 10^5$\,K or so, with this decrease happening at somewhat higher temperatures for more ionized species. The combination of the trend in the collision rate and the fractional abundance makes $\sim 10^5$K the sweet spot for emission in UV transitions. This figure also shows that at lower temperatures, photoionization by the ionizing background plays a more prominent role, with this evidenced by the bifurcation of the $n_e=10^{-2.0}$\cmminusthree (dashed curves) and $n_e=10^{-3.0}$\cmminusthree (solid). At higher temperatures, the ionization is determined by collisions, which in equilibrium with recombinations is independent of the density, and the density curves converge.

The emission estimates using the three previous equations in this paper were checked against {\sf Cloudy} calculations, which encode a more detailed atomic physics model.  This check shows excellent agreement over the regime where collisional emissions dominate and the ion is not highly ionized. This comparison confirms that there is not a cascade from a more excited state that sources a significant fraction of a line's photons, a process which our calculations would not be accounting for.  It would be more likely for such a cascade to be important for ions with hydrogen-like structure in which there is a pile up of $(n, \ell, m)$ states at $\sim 1$Ry above the ground state energy. This is not the case for the metal lines that are the most important emitters.  The trait of being important emission lines requires their ions to have low-lying UV and optical transitions to the ground state.

For the relevant atomic data for lines not listed in this paper, consult the CHIANTI Atomic Database to collect effective collision strengths from the .scups file and the Einstein A coefficients, needed to determine the branching ratio, from the .wgfa file associated with each species \citep{1997A&AS..125..149D, 2021ApJ...909...38D}. Effective collision strengths from CHIANTI are scaled according to the procedures summarized in the appendix of \citet{1992A&A...254..436B}.

\end{document}